\documentclass [useAMS]{mn2e}
\usepackage {graphicx}

\begin{document}

\title{A TOY MODEL FOR MAGNETIC CONNECTION IN BLACK-HOLE ACCRETION
DISC}
\author[]{Ding-Xiong Wang$^{*}$, Yong-Chun Ye, Yang Li and Dong-Mei Liu \\
$$ Department of Physics, Huazhong University of Science and Technology, Wuhan,430074,China \\
$^*$ Send offprint requests to: D.-X. Wang (dxwang@hust.edu.cn) }
\maketitle

\begin{abstract}

A toy model for magnetic connection in black hole (BH) accretion
disc is discussed based on a poloidal magnetic field generated by a
single electric current flowing around a Kerr black hole in the
equatorial plane. We discuss the effects of the coexistence of two
kinds of magnetic connection (MC) arising respectively from
(\ref{eq1}) the closed field lines connecting the BH horizon with
the disc (henceforth MCHD), and (\ref{eq2}) the closed field lines
connecting the plunging region with the disc (henceforth MCPD). The
magnetic field configuration is constrained by conservation of
magnetic flux and a criterion of the screw instability of the
magnetic field. Two parameters $\lambda $ and $\alpha _m $ are
introduced to describe our model instead of resolving the
complicated MHD equations. Compared with MCHD, energy and angular
momentum of the plunging particles are extracted via MCPD more
effectively, provided that the BH spin is not very high. It turns
out that negative energy can be delivered to the BH by the plunging
particles without violating the second law of BH thermodynamics,
however it cannot be realized via MCPD in a stable way.

\end{abstract}

\begin{keywords}
accretion, accretion disc --- black hole physics --- magnetic fields
\end{keywords}

\section{INTRODUCTION}

Recently, much attention has been paid to the magnetic connection
(MC) of a rotating black hole (BH) with its surrounding accretion
disc (Blandford 1999; Li 2000a; Wang, Xiao {\&} Lei 2002, Wang et
al. 2003, hereafter W02 and W03, respectively). The MC process can
be regarded as one of the variants of the Blandford-Znajek (BZ)
process proposed near three decades ago (Blandford {\&} Znajek
1977), which involves the closed magnetic field lines connecting a
BH with its surrounding disc. This mechanism has been used to
explain a very steep emissivity in the inner region of the disc,
which is consistent with the \textit{XMM-Newton} observation of the
nearby bright Seyfert1 galaxy MCG-6-30-15 (Wilms 2001; Li 2002a;
W03).

Blandford (2002) described several electromagnetic ways to extract
the rotational energy associated with a spinning BH, in which the MC
between the plunging region and the disc (hereafter MCPD) is
included. However, MCPD has not been discussed in detail in the
previous work, and the origin of these magnetic field configurations
remains unclear.

Recently, Li (2002b, hereafter L02) discussed the MC between the BH
horizon and the disc (hereafter MCHD) by assuming a toroidal
electric current $I$ flowing on a circle of $r = {r}'$ in the
equatorial plane of a Kerr BH, i.e.,

\begin{equation}
\label{eq1} J^a = \frac{I}{r}\left( {\Delta \mathord{\left/
{\vphantom {\Delta A}} \right. \kern-\nulldelimiterspace} A}
\right)^{1 / 2}\left( {\partial \mathord{\left/ {\vphantom {\partial
{\partial \phi }}} \right. \kern-\nulldelimiterspace} {\partial \phi
}} \right)^a\delta (r - {r}')\delta (\cos \theta ).
\end{equation}

\noindent The metric around a Kerr BH is given in Boyer-Lindquist
coordinates (MacDonald and Thorne 1982, hereafter MT82)

\begin{equation}
\label{eq2} \begin{array}{l}ds^2 = - \left( {1 - \frac{2Mr}{\rho
^2}} \right)dt^2 \\ \quad\quad - \frac{4aMr\sin ^2\theta }{\rho
^2}dtd\varphi + \frac{\rho ^2}{\Delta }dr^2\mbox{ + }\rho ^2d\theta
^2 + \frac{A\sin ^2\theta }{\rho ^2}d\varphi ^2,\end{array}
\end{equation}

\noindent where the concerned Kerr metric parameters are defined as

\begin{equation}
\label{eq3} \left\{ {\begin{array}{l}
 A = \left( {r^2 + a^2} \right)^2 - a^2\Delta \sin ^2\theta ,\mbox{ }\rho ^2
= r^2 + a^2\cos ^2\theta ,\mbox{ } \\
 \Delta = r^2 + a^2 - 2Mr,\mbox{ }\varpi = \left( {A \mathord{\left/
{\vphantom {A {\rho ^2}}} \right. \kern-\nulldelimiterspace} {\rho
^2}} \right)^{1 \mathord{\left/ {\vphantom {1 2}} \right.
\kern-\nulldelimiterspace} 2}\sin \theta , \\
 \alpha = \left( {{\rho ^2\Delta } \mathord{\left/ {\vphantom {{\rho
^2\Delta } A}} \right. \kern-\nulldelimiterspace} A} \right)^{1
\mathord{\left/ {\vphantom {1 2}} \right. \kern-\nulldelimiterspace} 2}, \quad \omega = 2aMr / A\\
 \end{array}} \right.
\end{equation}

\noindent In equation (\ref{eq3}) $a$ is the specific angular
momentum of the BH, which is related to the BH mass $M$, angular
momentum $J$ and spin $a_ * $ by $a \equiv J \mathord{\left/
{\vphantom {J M}} \right. \kern-\nulldelimiterspace} M = a_ * M$.

Not long ago, Li (2000b, hereafter L00) discussed a scenario of
extracting energy from a Kerr BH through the plunging region, where
the open magnetic field lines connect plasma particles with remote
loads. It is argued in L00 that the energy extracted from the
particles can be so large that the particles have negative energy as
they fall into the BH, if the magnetic field is strong enough.

Motivated by the above works, we propose a scenario of extracting
energy from a Kerr BH based on the magnetic field configuration
arising from a toroidal electric current $I$ flowing at the
innermost stable circular orbit (ISCO). The magnetic field
configuration contains both MCHD and MCPD, being constrained by the
conservation of magnetic flux and a criterion of the screw
instability of the magnetic field. It is shown that MCPD is more
important than MCHD in transferring energy and angular momentum to
the disc for low BH spins. It turns out that negative energy can be
delivered to the BH by the plunging particles without violating the
second law of BH thermodynamics, however it cannot be realized via
MCPD in a stable way.

This paper is organized as follows. In $\S$ 2 we present a
description of our model, in which MCHD and MCPD coexist in a
large-scale magnetic field generated by a toroidal electric current
flowing at ISCO. In $\S$ 3 we calculate and compare the powers and
torques in MCHD and MCPD. In $\S$ 4 the condition of negative energy
of plunging particles is studied in based on some reasonable
constraints: (i) a positive accretion rate, (ii) an increasing BH
entropy, (iii) a stationary accretion and (iv) a stable energy
extraction for the plunging particles. In $\S$ 5 we compare the
efficiencies of releasing energy arising from disc accretion, MCHD
and MCPD. Finally, in $\S$ 6, we summarize our main results, and
discuss some issues related to our model. Throughout this paper the
geometric units $G = c = 1$ are used.

\section{DESCRIPTION OF OUR MODEL }

\subsection{Magnetic field configuration produced by a toroidal
current}

It is assumed that the magnetic field in the neighborhood of the BH
and its surrounding disc is ideally conducting and force-free, and
the magnetic field, electric field and other quantities are measured
by zero-angular-momentum observers defined by Bardeen, Press {\&}
Teukolsky (1972). The magnetic field configuration produced by a
toroidal current flowing at the inner edge of the disc as shown in
Figure 1, which contains both MCHD and MCPD. In this paper we assume
that the toroidal current flows at ISCO, being represented by the
symbol ``$ \otimes $''. The quantities $r_H $ and $r_{ms} $ are
respectively the radii of the BH horizon and ISCO, and they read
(Novikov {\&} Thorne 1973)

\begin{equation}
\label{eq4} r_{_H }= M\left( {1 + \sqrt {1 - a_ * ^2 } } \right),
\end{equation}

\begin{equation}
\label{eq5} \left\{ {\begin{array}{l}
 r_{ms} \equiv M\chi _{ms}^2 , \\
 \chi _{ms} = \left\{ {3 + A_2 \pm \left[ {\left( {3 - A_1 } \right)\left(
{3 + A_1 + 2A_2 } \right)} \right]^{1 / 2}} \right\}^{1 / 2},\mbox{ } \\
 A_1 = 1 + \left( {1 - a_ * ^2 } \right)^{1 / 3}\left[ {\left( {1 + a_ * }
\right)^{1 / 3} + \left( {1 - a_ * } \right)^{1 / 3}} \right], \\
 A_2 = \left( {3a_ * ^2 + A_1^2 } \right)^{1 / 2}. \\
 \end{array}} \right.
\end{equation}

\noindent The quantity $r_e $ is the radius of the infinite redshift
surface and it reads

\begin{equation}
\label{eq6} r_e \left( \theta \right) = M\left( {1 + \sqrt {1 - a_ *
^2 \cos ^2\theta } } \right).
\end{equation}

\begin{figure}
\vspace{0.5cm}
\begin{center}
\includegraphics[width=6cm]{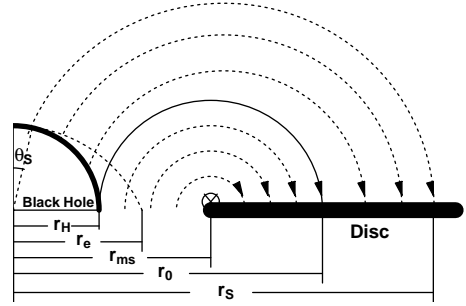}
\caption{Magnetic field configuration corresponding to MCHD and
MCPD, where the symbol ``$ \otimes $'' represents a toroidal current
at the inner edge of the disc.} \label{fig1}
\end{center}
\end{figure}

\noindent The radius $r_e $ increases with the polar angle $\theta
$, attaining its maximum at the equatorial plane, i.e., $r_e \left(
{\pi \mathord{\left/ {\vphantom {\pi 2}} \right.
\kern-\nulldelimiterspace} 2} \right) = 2M$. The range of the
ergosphere is given by

\begin{equation}
\label{eq7} {r}_{_H} < r < r_e \left( \theta \right),
\end{equation}

\noindent and the range of the plunging region is

\begin{equation}
\label{eq8} {r}_{_H} < r < r_{ms} .
\end{equation}

The magnetic flux through a surface bounded by a circle with $r =
$const and $\theta = $const is

\begin{equation}
\label{eq9} \Psi \left( {r,\theta ;{r}'} \right) = 2\pi A_\phi
\left( {r,\theta ;{r}'} \right),
\end{equation}

\noindent where $A_\phi $ is the toroidal component of the vector
potential produced by the toroidal electric current $I$ at $r =
{r}'$ in the equatorial plane of the Kerr BH. Defining $\tilde {r}
\equiv r \mathord{\left/ {\vphantom {r M}} \right.
\kern-\nulldelimiterspace} M$ and $B_0 \equiv {2I} \mathord{\left/
{\vphantom {{2I} M}} \right. \kern-\nulldelimiterspace} M$, we have
$\Psi \propto A_\varphi \propto MI = {B_0 M^2} \mathord{\left/
{\vphantom {{B_0 M^2} 2}} \right. \kern-\nulldelimiterspace} 2$ and

\begin{equation}
\label{eq10} \tilde {\Psi }\left( {\tilde {r},\theta ;\tilde {{r}'}}
\right) = {2\pi A_\varphi \left( {r,\theta ;{r}'} \right)}
\mathord{\left/ {\vphantom {{2\pi A_\varphi \left( {r,\theta ;{r}'}
\right)} {\left( {B_0 M^2} \right)}}} \right.
\kern-\nulldelimiterspace} {\left( {B_0 M^2} \right)}.
\end{equation}

Equation (\ref{eq10}) provides a mapping relation between any two
circles $\left( {\tilde {r}_1 ,\theta _1 } \right)$ and $\left(
{\tilde {r}_2 ,\theta _2 } \right)$ embedded in the same magnetic
surface as follows,

\begin{equation}
\label{eq11} \tilde {\Psi }\left( {\tilde {r}_1 ,\theta _1 ;\tilde
{{r}'}} \right) = \tilde {\Psi }\left( {\tilde {r}_2 ,\theta _2
;\tilde {{r}'}} \right).
\end{equation}

Znajek (1978) and Linet (1979) derived the relation between the
vector potential $A_\varphi \left( {r,\theta ;{r}'} \right)$ and the
current $I$ as follows.


\begin{equation}
\label{eq12}\begin{array}{l} A_\varphi = 2\sum\limits_{l = 1}^\infty
\{ {{\alpha _l^r \left[ {ra\sin ^2\theta \frac{\Delta }{\rho ^2}F_1
- a\sin ^2\theta \cos \theta F_2 } \right]}  } \\ \quad\quad +
\alpha _l^i \left[  - a^2\sin ^2\theta \cos \theta \frac{\Delta
}{\rho ^2}F_1 - r\sin ^2\theta F_2 + \frac{\Delta \sin ^2\theta
}{l\left( {l + 1} \right)}F_3  \right] \} \\ \quad\quad +
2\sum\limits_{l = 1}^\infty \{{ {\beta _l^r \left[ {ra\sin ^2\theta
\frac{\Delta }{\rho ^2}F_4 - a\sin ^2\theta \cos \theta F_5 }
\right]}} \\ \quad\quad + \beta _l^i \left[ { - a^2\sin ^2\theta
\cos \theta \frac{\Delta }{\rho ^2}F_4 }  - r\sin ^2\theta F_5  +
\frac{\Delta \sin ^2\theta }{l\left( {l + 1} \right)}F_6  \right]\}
\end{array}
\end{equation}

\noindent where in equation (\ref{eq12}) the functions $F_1 $ ---
$F_6 $ are defined by

\begin{equation}
\label{eq13} \left\{ {\begin{array}{l}
 F_1 \equiv \frac{1}{\sqrt {M^2 - a^2} }{P}'_l \left( u \right)P_l \left(
{\cos \theta } \right),\\ F_2 \equiv \frac{r^2 + a^2}{\rho ^2}P_l
\left( u \right){P}'_l \left( {\cos \theta } \right)\mbox{,} \\
 F_3 \equiv \frac{1}{\sqrt {M^2 - a^2} }{P}'_l \left( u \right){P}'_l \left(
{\cos \theta } \right), \\ F_4 \equiv \frac{1}{\sqrt {M^2 - a^2}
}{Q}'_l
\left( u \right)Q_l \left( {\cos \theta } \right), \\
 F_5 \equiv \frac{r^2 + a^2}{\rho ^2}Q_l \left( u \right){Q}'_l \left( {\cos
\theta } \right)\\ F_6 \equiv \frac{1}{\sqrt {M^2 - a^2} }{Q}'_l
\left( u \right){Q}'_l \left( {\cos \theta } \right)\mbox{. } \\
 \end{array}} \right.
\end{equation}

In equation (\ref{eq13}) we have $u \equiv {\left( {r - M} \right)}
\mathord{\left/ {\vphantom {{\left( {r - M} \right)} {\sqrt {M^2 -
a^2} }}} \right. \kern-\nulldelimiterspace} {\sqrt {M^2 - a^2} }$,
$P_l \left( z \right)$ and $Q_l \left( z \right)$ are Legendre
functions with ${P}'_l \left( z \right) \equiv {dP_l \left( z
\right)} \mathord{\left/ {\vphantom {{dP_l \left( z \right)} {dz}}}
\right. \kern-\nulldelimiterspace} {dz}$ and ${Q}'_l \left( z
\right) \equiv {dQ_l \left( z \right)} \mathord{\left/ {\vphantom
{{dQ_l \left( z \right)} {dz}}} \right. \kern-\nulldelimiterspace}
{dz}$. The coefficients $\alpha _l^r $, $\alpha _l^i $, $\beta _l^r
$ and $\beta _l^i $ are defined as follows:

(1) for $r < {r}'$, $\beta _l^r = \beta _l^i = 0$ for all $l$; but


\begin{equation}
\label{eq14} \alpha _l^r \equiv \frac{\pi \left( {2l + 1}
\right)I}{l\left( {l + 1} \right)\left( {M^2 - a^2} \right)}\left(
{\frac{{\Delta }'}{{A}'}} \right)^{1 \mathord{\left/ {\vphantom {1
2}} \right. \kern-\nulldelimiterspace} 2}{\Delta }'aP_l \left( 0
\right){Q}'_l \left( {u}' \right),
\end{equation}


\begin{equation}
\label{eq15}
\begin{array}{l}
 \alpha _l^i \equiv \frac{\pi \left( {2l + 1} \right)I}{l\left( {l + 1}
\right)\left( {M^2 - a^2} \right)}\left( {\frac{{\Delta }'}{{A}'}}
\right)^{1 \mathord{\left/ {\vphantom {1 2}} \right.
\kern-\nulldelimiterspace} 2} \\
 \quad\quad \times [ { - \left( {{r}'^2 + a^2} \right){P}'_l \left( 0
\right)Q_l \left( {u}' \right)}\\ \quad\quad +  \frac{{r}'{\Delta
}'}{l\left( {l + 1} \right)}\frac{1}{\sqrt {M^2 - a^2} }{P}'_l
\left( 0 \right){Q}'_l \left(
{u}' \right) ]; \\
 \end{array}
\end{equation}

(2) for $r > {r}'$, $\alpha _l^r = \alpha _l^i = 0$ for all $l$; but


\begin{equation}
\label{eq16} \beta _l^r \equiv \frac{\pi \left( {2l + 1}
\right)I}{l\left( {l + 1} \right)\left( {M^2 - a^2} \right)}\left(
{\frac{{\Delta }'}{{A}'}} \right)^{1 \mathord{\left/ {\vphantom {1
2}} \right. \kern-\nulldelimiterspace} 2}{\Delta }'aP_l \left( 0
\right){P}'_l \left( {u}' \right),
\end{equation}


\begin{equation}
\label{eq17}
\begin{array}{l}
 \beta _l^i \equiv \frac{\pi \left( {2l + 1} \right)I}{l\left( {l + 1}
\right)\left( {M^2 - a^2} \right)}\left( {\frac{{\Delta }'}{{A}'}}
\right)^{1 \mathord{\left/ {\vphantom {1 2}} \right.
\kern-\nulldelimiterspace} 2} \\
\quad\quad \times [ { - \left( {{r}'^2 + a^2} \right){P}'_l \left( 0
\right)P_l \left( {u}' \right)} \\ \quad\quad + \frac{{r}'{\Delta
}'}{l\left( {l + 1} \right)}\frac{1}{\sqrt {M^2 - a^2} }{P}'_l
\left( 0 \right){P}'_l \left(
{u}' \right)], \\
 \end{array}
\end{equation}

\noindent where in equations (\ref{eq14})---(\ref{eq17}) we have
${\Delta }' \equiv \Delta \left( {r = {r}'} \right)$ and ${A}'
\equiv A\left( {r = {r}',\mbox{ }\theta = \pi \mathord{\left/
{\vphantom {\pi 2}} \right. \kern-\nulldelimiterspace} 2} \right)$.

\subsection{Mapping relations for MCHD and MCPD}

For the toroidal current located at $r_{ms} $ the mapping relation
between the circle at the BH horizon with the longitudinal angle
$\theta $ and the circle at the thin disc with the radius $r$ can be
written as

\begin{equation}
\label{eq18} \tilde {\Psi }\left( {\tilde {r},\pi \mathord{\left/
{\vphantom {\pi 2}} \right. \kern-\nulldelimiterspace} 2;\tilde
{r}_{ms} } \right) = \tilde {\Psi }\left( {\tilde {r}_{_H} ,\theta
;\tilde {r}_{ms} } \right).
\end{equation}

It is assumed that the toroidal current located at $r_{ms} $ has a
very small circular section of radius ${r}'_\varepsilon =
\varepsilon r_{ms} $, so that the poloidal magnetic field is
proportional to the radius ${r}'$ for $0 < {r}' < {r}'_\varepsilon
$. In this way the infinite magnetic field is avoided as $r $close
to $r_{ms} $, keeping magnetic field configuration outside
${r}'_\varepsilon $ unchanged. In this paper we take $\varepsilon =
10^{ - 4}$ in calculations (The influence of the choice of
$\varepsilon $ is discussed in $\S$ 6).

The innermost magnetic surface for MCHD is defined as a surface
connecting the horizon at $\theta = \pi \mathord{\left/ {\vphantom
{\pi 2}} \right. \kern-\nulldelimiterspace} 2$ with the disc at
$\tilde {r}_0 \ge \tilde {r}_{ms} $, and it can be determined by

\begin{equation}
\label{eq19} \tilde {\Psi }\left( {\tilde {r}_0 ,\pi \mathord{\left/
{\vphantom {\pi 2}} \right. \kern-\nulldelimiterspace} 2;\tilde
{r}_{ms} } \right) = \tilde {\Psi }\left( {\tilde {r}_{_H} ,\pi
\mathord{\left/ {\vphantom {\pi 2}} \right.
\kern-\nulldelimiterspace} 2;\tilde {r}_{ms} } \right),
\end{equation}

\noindent where $\tilde {r}_0 \equiv {r_0 } \mathord{\left/
{\vphantom {{r_0 } M}} \right. \kern-\nulldelimiterspace} M$ is the
radial coordinate of the disc plane intersected with the innermost
magnetic surface in MCHD as shown by the solid field line in Figure
1.

Equation (\ref{eq11}) can be applied to MCPD by taking $\theta _1 =
\theta _2 = \pi \mathord{\left/ {\vphantom {\pi 2}} \right.
\kern-\nulldelimiterspace} 2$. The radial coordinate $\tilde
{r}_{_{PL}} $ in the plunging region and $\tilde {r}$ in the disc
are related by

\begin{equation}
\label{eq20} \left\{ {\begin{array}{l}
 \tilde {\Psi }\left( {\tilde {r}_{_{PL}} ,\pi \mathord{\left/ {\vphantom {\pi
2}} \right. \kern-\nulldelimiterspace} 2;\tilde {r}_{ms} } \right) =
\tilde {\Psi }\left( {\tilde {r},\pi \mathord{\left/ {\vphantom {\pi
2}} \right.
\kern-\nulldelimiterspace} 2;\tilde {r}_{ms} } \right), \\
 \tilde {r}_{_H} < \tilde {r}_{_{PL}} < \tilde {r}_{ms} , \\
 \tilde {r}_{ms} < \tilde {r} < \tilde {r}_0 . \\
 \end{array}} \right.
\end{equation}

Equation (\ref{eq20}) is the mapping relation for MCPD, which
relates the radius $r_{_{PL}} $ in the plunging region to the radius
$r$ at the disc.

The magnetic field configuration can be constrained by the screw
instability, which will occur if the toroidal magnetic field becomes
so strong that the magnetic field line turns around itself about
once (Kadomtsev 1966; Bateman 1978). Recently, Wang et al. (2004)
argued that the screw instability would occur in a magnetized
accretion disc, and the criterion can be expressed as

\begin{equation}
\label{eq21} \left( {{2\pi \varpi _{_D} } \mathord{\left/ {\vphantom
{{2\pi \varpi _{_D} } L}} \right. \kern-\nulldelimiterspace} L}
\right)B_D^P / B_D^T \le 1,
\end{equation}

\noindent where $B_D^p $ and $B_D^T $ are respectively the poloidal
and toroidal components of the magnetic field on the disc, and $L$
is the poloidal length of the closed field line connecting the BH
with the disc. The symbol $\varpi _{_D} $ is the cylindrical radius
on the disc and it reads

\begin{equation}
\label{eq22}\begin{array}{l} \varpi _{_D} = {\sqrt {A_D } }
\mathord{\left/ {\vphantom {{\sqrt {A_D } } {\rho _{_D} }}} \right.
\kern-\nulldelimiterspace} {\rho _{_D} } \\ \quad\quad = \xi M\chi
_{ms}^2 \sqrt {1 + a_ * ^2 \xi ^{ - 2}\chi _{ms}^{ - 4} + 2a_ * ^2
\xi ^{ - 3}\chi _{ms}^{ - 6} } .\end{array}
\end{equation}

\noindent where $\xi \equiv r \mathord{\left/ {\vphantom {r {r_{ms}
}}} \right. \kern-\nulldelimiterspace} {r_{ms} }$ is a dimensionless
radial parameter. The poloidal component of the magnetic field
$B_D^p $ can be written as

\begin{equation}
\label{eq23} B_D^P = - \frac{\left( {{\Delta _D } \mathord{\left/
{\vphantom {{\Delta _D } {A_D }}} \right. \kern-\nulldelimiterspace}
{A_D }} \right)^{1 / 2}}{2\pi }\frac{d\Psi (r,\pi \mathord{\left/
{\vphantom {\pi {2;r_{ms} )}}} \right. \kern-\nulldelimiterspace}
{2;r_{ms} )}}{dr}.
\end{equation}

\noindent Equation (\ref{eq23}) is derived based on the calculation
on the magnetic flux between two adjacent magnetic surfaces, i.e.,

\begin{equation}
\label{eq24} d\Psi (r,\pi \mathord{\left/ {\vphantom {\pi {2;r_{ms}
)}}} \right. \kern-\nulldelimiterspace} {2;r_{ms} )} = - B_D^P 2\pi
\sqrt {g_{rr} g_{\varphi \varphi } } dr.
\end{equation}

In W02 the powers and torques in the BZ and MCHD processes are
derived based on an equivalent circuit consisting of a series of
loops, which correspond to a series of adjacent magnetic surfaces.
By using Ampere's law the toroidal magnetic field $B_D^T $ can be
expressed as

\begin{equation}
\label{eq25} B_D^T = {2I_{MC}^{HD} } \mathord{\left/ {\vphantom
{{2I_{MC}^{HD} } {\left( {\varpi _{_D} \alpha } \right)_{\theta =
\pi \mathord{\left/ {\vphantom {\pi 2}} \right.
\kern-\nulldelimiterspace} 2} }}} \right. \kern-\nulldelimiterspace}
{\left( {\varpi _{_D} \alpha } \right)_{\theta = \pi \mathord{\left/
{\vphantom {\pi 2}} \right. \kern-\nulldelimiterspace} 2} }.
\end{equation}

\noindent The electric current $I_{MC}^{HD} $ flowing in each loop
of the equivalent circuit for MCHD is given by

\begin{equation}
\label{eq26} I_{MC}^{HD} = \frac{\Delta \varepsilon _{_H} + \Delta
\varepsilon _{_D} }{\Delta Z_H + \Delta Z_D }.
\end{equation}

The resistance $\Delta Z_D $ in equation (\ref{eq26}) can be
neglected due to the perfect conductivity of the disc plasmas. The
quantities $\Delta \varepsilon _{_H} $ and $\Delta \varepsilon _{_D}
$ are electromotive forces due to the rotation of the BH and the
disc, respectively, and they read

\begin{equation}
\label{eq27} \left\{ {\begin{array}{l}
 \Delta \varepsilon _{_H} = \left[ {{\left( {\Delta \Psi } \right)_H }
\mathord{\left/ {\vphantom {{\left( {\Delta \Psi } \right)_H } {2\pi
}}}
\right. \kern-\nulldelimiterspace} {2\pi }} \right]\Omega _H , \\
 \Delta \varepsilon _{_D} = - \left[ {{\left( {\Delta \Psi } \right)_D }
\mathord{\left/ {\vphantom {{\left( {\Delta \Psi } \right)_D } {2\pi
}}}
\right. \kern-\nulldelimiterspace} {2\pi }} \right]\Omega _D , \\
 \end{array}} \right.
\end{equation}

\noindent where $\left( {\Delta \Psi } \right)_H $ and $\left(
{\Delta \Psi } \right)_D $ are the fluxes sandwiched between two
adjacent magnetic surfaces threading the BH horizon and the disc,
respectively. The minus sign in the expression of $\Delta
\varepsilon _{_D} $ arises from the direction of the flux. We have
$\left( {\Delta \Psi } \right)_H = \left( {\Delta \Psi } \right)_D $
and the two can be calculated by using equation (\ref{eq24}). The
quantities $\Omega _H $ and $\Omega _D $ are respectively the
angular velocity of the BH horizon and that of the disc, and they
read

\begin{equation}
\label{eq28} \Omega _H = \frac{a_ * }{2M\tilde {r}_{_H} }, \quad
\Omega _D = \frac{1}{M\left( {\tilde {r}^{3 \mathord{\left/
{\vphantom {3 2}} \right. \kern-\nulldelimiterspace} 2} + a_ * }
\right)}.
\end{equation}

Incorporating equations (\ref{eq21})---(\ref{eq28}), we have the
criterion of the screw instability as follows,

\begin{equation}
\label{eq29} \frac{\left( {{\Delta _D } \mathord{\left/ {\vphantom
{{\Delta _D } {A_D }}} \right. \kern-\nulldelimiterspace} {A_D }}
\right)^{1 / 2}\left( {{ - dZ_H } \mathord{\left/ {\vphantom {{ -
dZ_H } {dr}}} \right. \kern-\nulldelimiterspace} {dr}} \right)\left(
{\varpi _{_D} \alpha } \right)_{\theta = \pi \mathord{\left/
{\vphantom {\pi 2}} \right. \kern-\nulldelimiterspace} 2} }{2\left(
{\Omega _H - \Omega _D } \right)} \le L \mathord{\left/ {\vphantom
{L {\left( {2\pi \varpi _{_D} } \right)}}} \right.
\kern-\nulldelimiterspace} {\left( {2\pi \varpi _{_D} } \right)}.
\end{equation}

\noindent In equation (\ref{eq29}) $dZ_H $ is the resistance of the
BH horizon between the two adjacent magnetic surfaces, and we have

\begin{equation}
\label{eq30} {dZ_H } \mathord{\left/ {\vphantom {{dZ_H } {dr}}}
\right. \kern-\nulldelimiterspace} {dr} = \left( {{dZ_H }
\mathord{\left/ {\vphantom {{dZ_H } {d\theta }}} \right.
\kern-\nulldelimiterspace} {d\theta }} \right)\left( {{d\theta }
\mathord{\left/ {\vphantom {{d\theta } {dr}}} \right.
\kern-\nulldelimiterspace} {dr}} \right),
\end{equation}

\noindent where ${d\theta } \mathord{\left/ {\vphantom {{d\theta }
{dr}}} \right. \kern-\nulldelimiterspace} {dr} < 0$ can be
calculated by using the mapping relation (\ref{eq18}).

Incorporating equations (\ref{eq21})---(\ref{eq30}), we can derived
the critical radius of the screw instability, $\tilde {r}_{_S}
\equiv {r_{_S} } \mathord{\left/ {\vphantom {{r_{_S} } M}} \right.
\kern-\nulldelimiterspace} M$. As was demonstrated by Uzdensky
(2005), the pressure of the toroidal magnetic field wound up by the
rotating BH cannot be compensated in the outer regions of the
magnetosphere. Thus we regard the radius $r_{_S} $ as the outer
boundary of MCHD region as shown in Figure 1. The critical radius
$r_{_S} $ is related to the angle $\theta _S $ by the mapping
relation (\ref{eq18}). More specifically, the mapping relation for
MCHD can be expressed as

\begin{equation}
\label{eq31} \left\{ {\begin{array}{l}
 \Psi \left( {{r}_{_H} ,\theta ;r_{ms} } \right) = \Psi \left( {r,\pi
\mathord{\left/ {\vphantom {\pi 2}} \right.
\kern-\nulldelimiterspace}
2;r_{ms} } \right), \\
 \theta _S < \theta < \pi \mathord{\left/ {\vphantom {\pi {2,}}} \right.
\kern-\nulldelimiterspace} {2,} \\
 r_0 < r < r_{_S} . \\
 \end{array}} \right.
\end{equation}

Thus the magnetic field configuration produced by the toroidal
current is constrained by the conservation of magnetic flux and the
criterion of the screw instability. By using equations (\ref{eq4}),
(\ref{eq5}), (\ref{eq6}), (\ref{eq19}) and (\ref{eq29}) we have the
curves of the radii $\tilde {r}_{_H} $, $\tilde {r}_{ms} $, $\tilde
{r}_e $, $\tilde {r}_0 $ and $\tilde {r}_{_S} $ versus $a_ * $ as
shown in Figure 2.

\begin{figure}
\vspace{0.5cm}
\begin{center}
{\includegraphics[width=6cm]{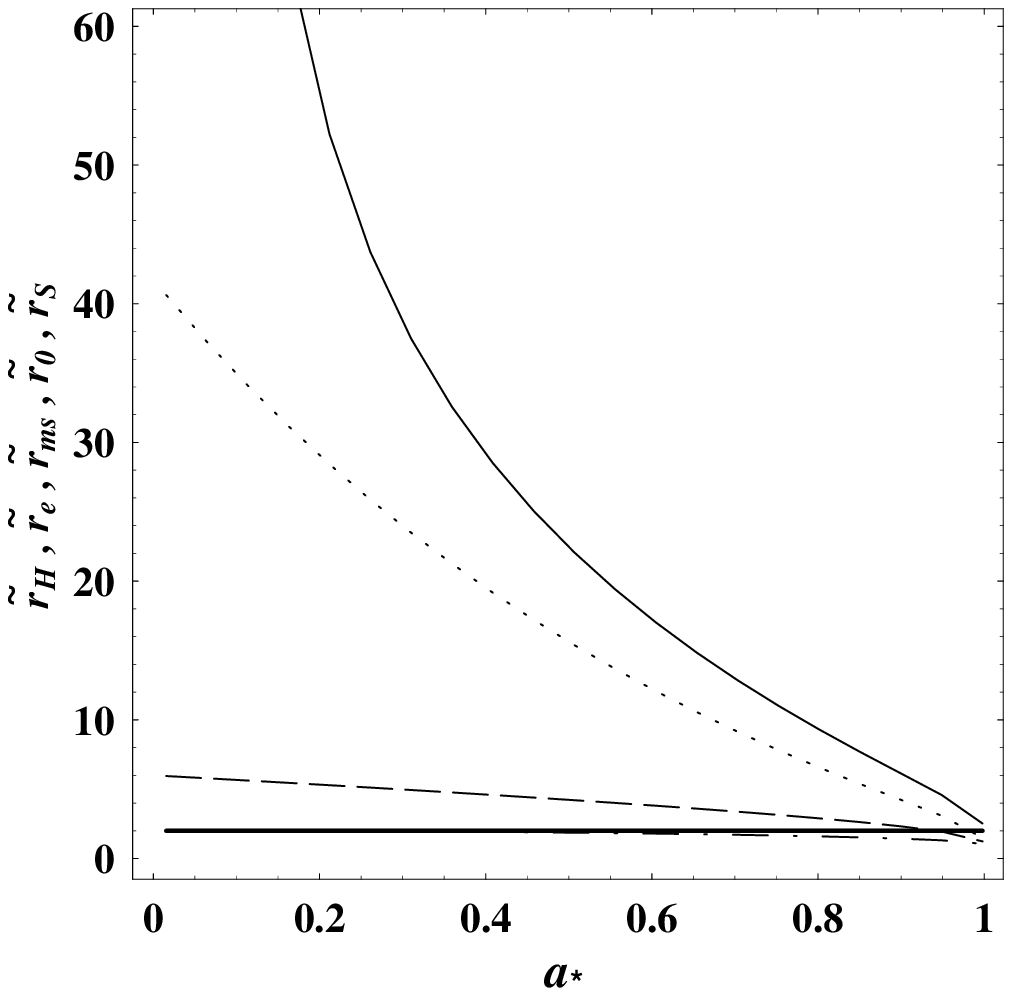}
 \centerline{(a)}
 \includegraphics[width=6cm]{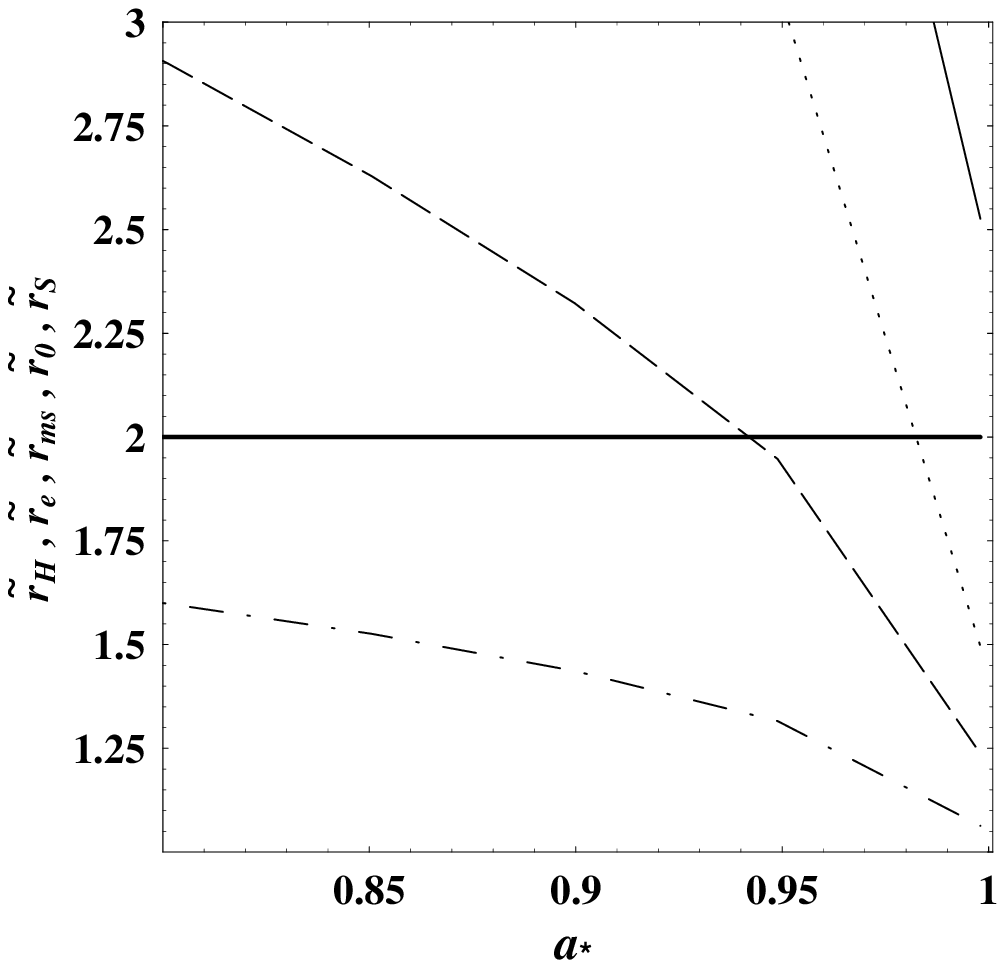}
 \centerline{(b)}}
 \caption{Curves of $\tilde {r}_{_H} $, $\tilde
{r}_{ms} $, $\tilde {r}_e $, $\tilde {r}_{_0} $ and $\tilde {r}_{_S}
$ versus $a_ * $ in dot-dashed, dashed, thick solid, dotted and thin
solid lines, respectively, for $0 < a_ * < 0.998$ (a) and $0.8 < a_
* < 0.998$ (b).}\label{fig2}
\end{center}
\end{figure}

It is shown that the radii $\tilde {r}_{_H} $, $\tilde {r}_{ms} $,
$\tilde {r}_0 $ and $\tilde {r}_{_S} $ decrease monotonically with
the increasing $a_ * $, while the radius $\tilde {r}_e $ remaining a
constant ``2'' in the equatorial plane. It is easy to check from
equations (\ref{eq5}) and (\ref{eq6}) that $\tilde {r}_{ms} = \tilde
{r}_e $ holds for $a_ * = \sqrt {8 \mathord{\left/ {\vphantom {8 9}}
\right. \kern-\nulldelimiterspace} 9} = 0.9428$, beyond which the
inner edge of the disc will enter into the ergosphere.

\section{POWERS AND TORQUES IN MCHD AND MCPD}

The motion of the plunging particles is very different from that in
the disc. The velocity of the radial inflow in the disc is much
smaller than the Keplerian velocity, while the radial velocity of
plunging particles is comparable to the Keplerian velocity. The
motion of plunging particles becomes even more complicated in the
presence of large-scale magnetic fields. In this paper, instead of
resolving MHD equations in the plunging region, we introduce a
parameter $\lambda $ to describe the MCPD effect on the motion of
the plunging particles.

\subsection{Angular velocity of plunging particles}

It is reasonable to assume that the accreting matter mainly consists
of electrons and protons. Considering the toroidal motion of the
accreting electrons and protons in the plunging region and the
magnetic field configuration in Figure 1, we infer that the
electrons of negative charge and the protons of positive charge are
accelerated inwards and decelerated outwards by Lorentz force,
respectively. Thus a current is driven outwards in the plunging
region by Lorentz force. This situation is analogous to those in the
BZ process and in MCHD, where a current is driven in the
longitudinal direction at the horizon due to the BH rotation
relative to a poloidal large-scale magnetic field (MT82; W02; W03).
And we expect to derive the power and torque in MCPD by using an
equivalent circuit analogous to that given for MCHD in W02 and W03.

Starting from the Lagrangian of a test particle (Shapiro {\&}
Teukolsky 1983), we can derive the expression for the angular
velocity $\Omega _{PL} $ of the particle in the plunging region as
follows,

\begin{equation}
\label{eq32} \Omega _{PL} = \frac{1}{M}\frac{\left( {\tilde
{r}_{_{PL}} - 2} \right)\left( {{L_{ms} } \mathord{\left/ {\vphantom
{{L_{ms} } M}} \right. \kern-\nulldelimiterspace} M} \right) + 2a_ *
E_{ms} }{\left( {\tilde {r}_{_{PL}}^3 + a_ * ^2 \tilde {r}_{_{PL}} +
2a_ * ^2 } \right)E_{ms} - 2a_ * \left( {{L_{ms} } \mathord{\left/
{\vphantom {{L_{ms} } M}} \right. \kern-\nulldelimiterspace} M}
\right)},
\end{equation}

\noindent where $E_{ms} $ and $L_{ms} $ are the specific energy and
angular momentum of the accreting particles at ISCO, and they read

\begin{equation}
\label{eq33} E_{ms} = \frac{1 - 2\tilde {r}_{ms}^{ - 1} + a_ *
\tilde {r}_{ms}^{{ - 3} \mathord{\left/ {\vphantom {{ - 3} 2}}
\right. \kern-\nulldelimiterspace} 2} }{\left( {1 - 3\tilde
{r}_{ms}^{ - 1} + 2a_ * \tilde {r}_{ms}^{{ - 3} \mathord{\left/
{\vphantom {{ - 3} 2}} \right. \kern-\nulldelimiterspace} 2} }
\right)^{1 / 2}},
\end{equation}

\begin{equation}
\label{eq34} {L_{ms} } \mathord{\left/ {\vphantom {{L_{ms} } M}}
\right. \kern-\nulldelimiterspace} M = \frac{\tilde {r}_{ms}^{1
\mathord{\left/ {\vphantom {1 2}} \right. \kern-\nulldelimiterspace}
2} \left( {1 - 2a_ * \tilde {r}_{ms}^{{ - 3} \mathord{\left/
{\vphantom {{ - 3} 2}} \right. \kern-\nulldelimiterspace} 2} + a_ *
^2 \tilde {r}_{ms}^{ - 2} } \right)}{\left( {1 - 3\tilde {r}_{ms}^{
- 1} + 2a_ * \tilde {r}_{ms}^{{ - 3} \mathord{\left/ {\vphantom {{ -
3} 2}} \right. \kern-\nulldelimiterspace} 2} } \right)^{1 / 2}}.
\end{equation}

Considering the magnetic extraction of energy and angular momentum,
we suggest that the angular velocity, $\Omega _{PL}^m $, of the
plunging particle is related to $\Omega _{PL} $ by

\begin{equation}
\label{eq35} \Omega _{PL}^m = \left( {1 - \lambda \cos \tau }
\right)\Omega _{PL} ,
\end{equation}

\noindent where $\lambda $ is a parameter to adjust the ratio of
$\Omega _{PL}^m $ to $\Omega _{PL} $, of which the value range is $0
< \lambda < 1$. The parameter $\tau $ is defined as

\begin{equation}
\label{eq36} \tau \equiv \frac{\left( {2\tilde {r}_{_{PL}} - \tilde
{r}_{ms} - \tilde {r}_{_H} } \right)\pi \mathord{\left/ {\vphantom
{\pi 2}} \right. \kern-\nulldelimiterspace} 2}{\tilde {r}_{ms} -
\tilde {r}_{_H} }, \quad \tilde {r}_{_H} < \tilde {r}_{_{PL}} <
\tilde {r}_{ms} .
\end{equation}

Incorporating equations (\ref{eq32}), (\ref{eq35}) and (\ref{eq36}),
we find that the angular velocity $\Omega _{PL}^m $ is decelerated
at most to $\left( {1 - \lambda } \right)\Omega _{PL} $ at $\tilde
{r}_{_{PL}} = {\left( {\tilde {r}_{ms} + \tilde {r}_{_H} } \right)}
\mathord{\left/ {\vphantom {{\left( {\tilde {r}_{ms} + \tilde
{r}_{_H} } \right)} 2}} \right. \kern-\nulldelimiterspace} 2$, and
it connects smoothly with $\Omega _H $ and $\Omega _D $ at ${r}_{_H}
$ and $r_{ms} $, respectively.

For the given values of $\lambda $ the angular velocity $\Omega
_{PL}^m $ is a function of $a_ * $ and $\tilde {r}_{_{PL}} $, and we
have the curves of ${\Omega _{PL}^m } \mathord{\left/ {\vphantom
{{\Omega _{PL}^m } {\Omega _0 }}} \right. \kern-\nulldelimiterspace}
{\Omega _0 }$ varying with $\tilde {r}_{_{PL}} $ for the given BH
spins as shown in Figure 3.

\begin{figure*}
\begin{center}
{\includegraphics[width=5.0cm]{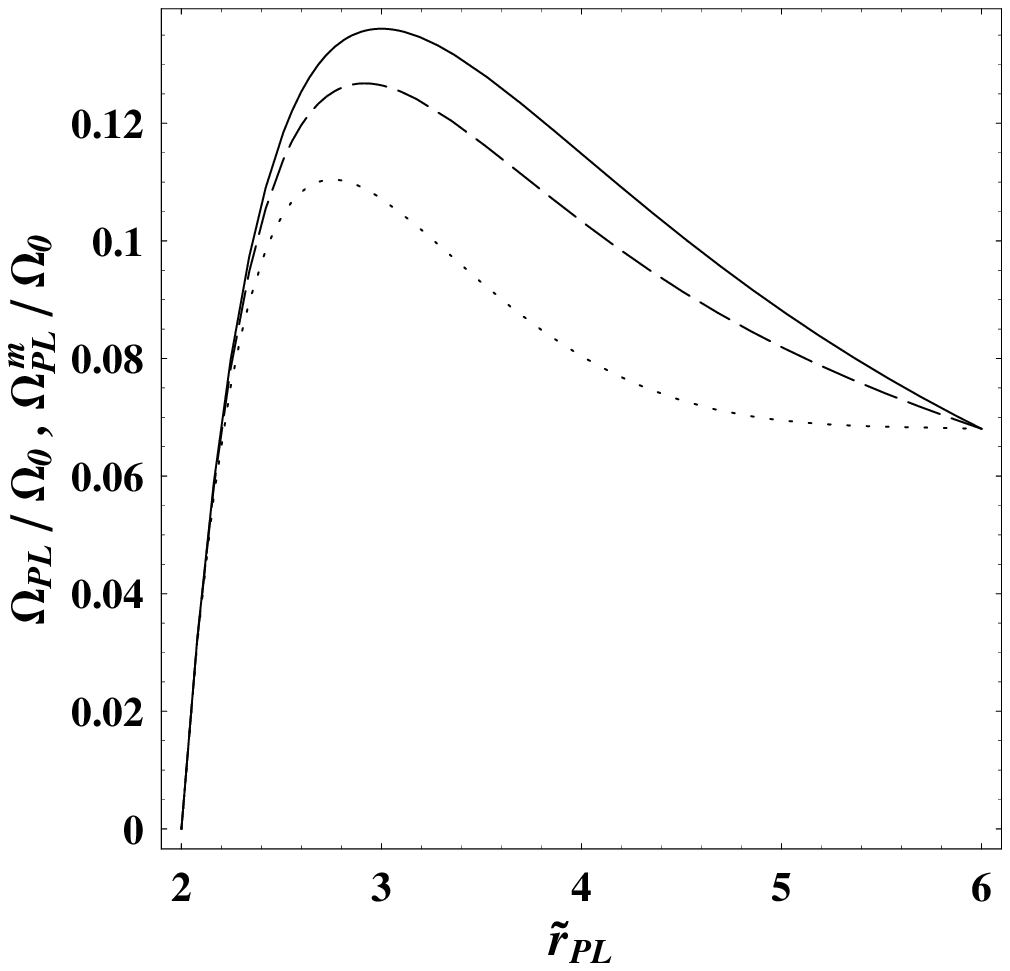} \hfill
\includegraphics[width=5.0cm]{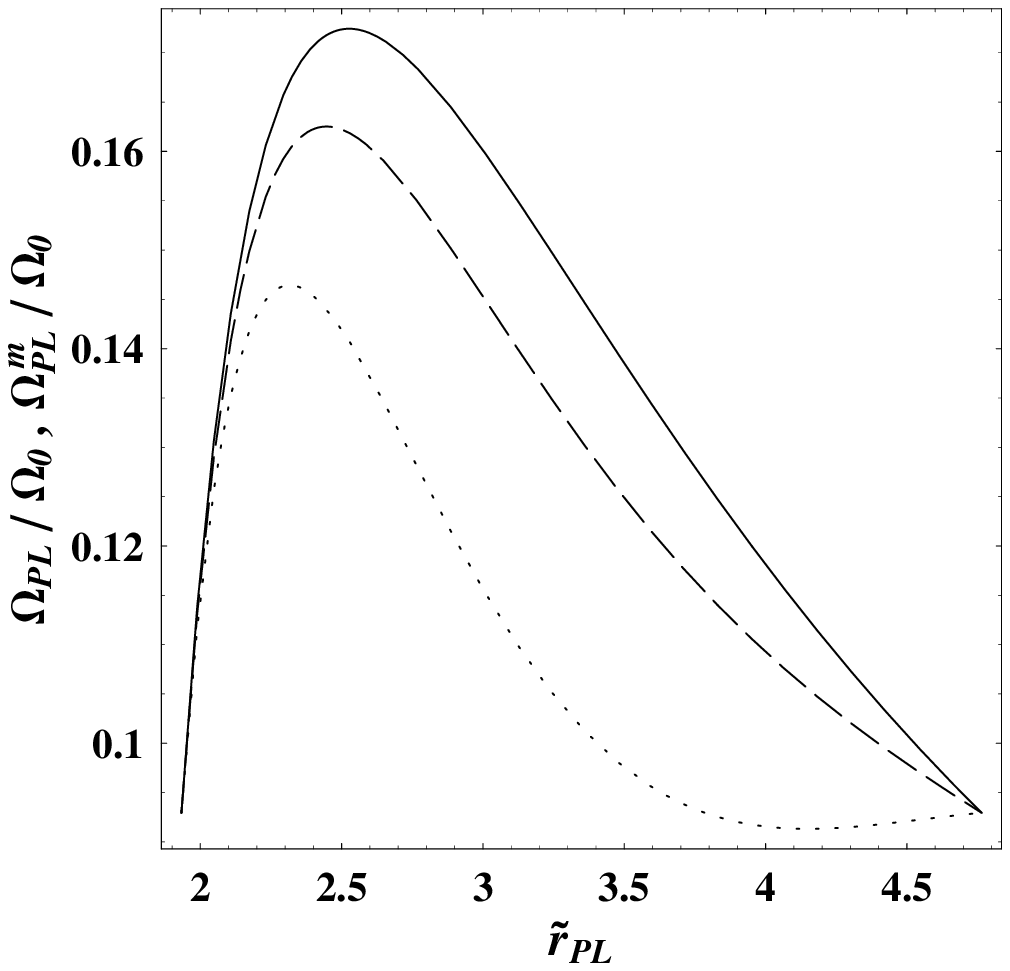} \hfill
\includegraphics[width=5.0cm]{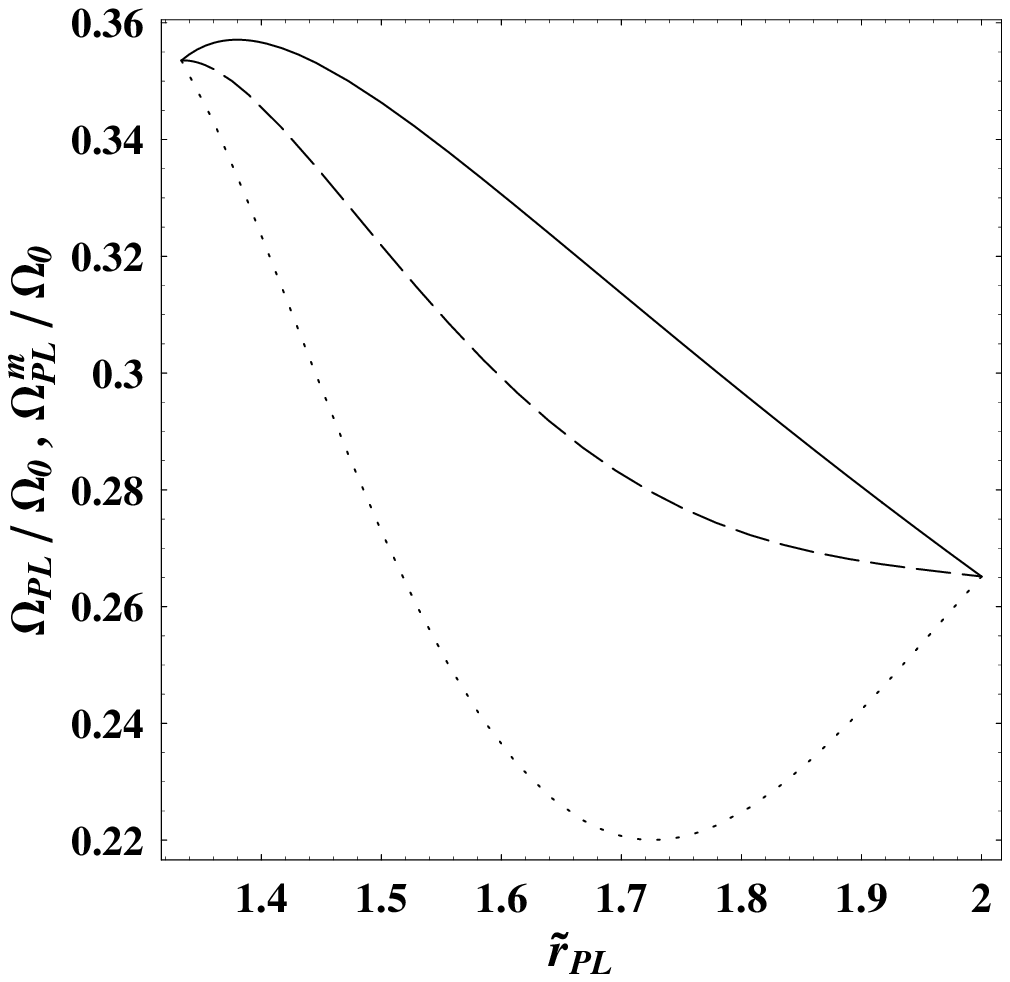}
\centerline{\hspace{0.6cm}(a)\hspace{6cm}(b)\hspace{6cm}(c)} }
\caption{The curves of $\Omega _{PL} / \Omega _0 $ and ${\Omega
_{PL}^m } / {\Omega _0 }$ versus $\tilde {r}_{_{PL}} $ with $a_ * =
0, 0.3594, 0.9428$ in (a), (b), and (c), respectively, where
${\Omega _{PL} } / {\Omega _0 }$ is plotted in solid line with
$\lambda $=0, and ${\Omega _{PL}^m } / {\Omega _0 }$ is plotted in
dashed and dotted lines with $\lambda $=0.1 and 0.3, respectively.
The value range of $\tilde {r}_{_{PL}} $ is $\tilde {r}_{_{H}} <
\tilde {r}_{_{PL}} < \tilde {r}_{ms} $ and $\Omega _0 \equiv 1 / M =
2.03\times 10^5 (rad/s) ( M / M_ {\odot} )^{ - 1}$} \label{fig3}
\end{center}
\end{figure*}

It is shown in Figure 3 that the profile of $\Omega _{PL}^m $ is
affected significantly by the parameter $\lambda $ at the middle of
the plunging region, where a less $\Omega _{PL}^m $ corresponds to a
larger $\lambda $. The value of $\lambda $ is crucial for MCPD in
transferring energy and angular momentum, and we shall discuss its
value range based on some physical considerations.

\subsection{Derivation of powers and torques in MCHD AND MCPD}

We can derive the expressions for power and torque in MCPD by using
an analogous equivalent circuit given in W02, and the electromotive
force due to the rotation of the plunging matter in the plunging
region is written as

\begin{equation}
\label{eq37} \Delta \varepsilon _{PL} = \left[ {{\left( {\Delta \Psi
} \right)_{PL} } \mathord{\left/ {\vphantom {{\left( {\Delta \Psi }
\right)_{PL} } {2\pi }}} \right. \kern-\nulldelimiterspace} {2\pi }}
\right]\Omega _{PL}^m ,
\end{equation}

\noindent where $\left( {\Delta \Psi } \right)_{PL} $ is the
magnetic flux threading the plunging region between the two adjacent
magnetic surfaces and it reads

\begin{equation}
\label{eq38} \begin{array}{l} \left( {\Delta \Psi } \right)_{PL} =
B_D^P 2\pi \sqrt {g_{rr} g_{\varphi \varphi } } \Delta r \\
\quad\quad\quad\quad = B_D^P M^22\pi \tilde {r}\Delta \tilde
{r}\sqrt {\frac{\tilde {r}^2 + a_ * ^2 + 2{a_ * ^2 } \mathord{\left/
{\vphantom {{a_
* ^2 } \tilde {r}}} \right. \kern-\nulldelimiterspace} \tilde {r}}{\tilde
{r}^2 + a_ * ^2 - 2\tilde {r}}} .\end{array}
\end{equation}

\noindent The electric current in each loop of the equivalent
circuit can be written as

\begin{equation}
\label{eq39} I_{MC}^{PD} = \frac{\Delta \varepsilon _{PL} + \Delta
\varepsilon _{_D} }{\Delta Z_{PL} },
\end{equation}

\noindent where $\Delta \varepsilon _{_D} $ and $\Delta \varepsilon
_{PL} $ are expressed by equations (\ref{eq27}) and (\ref{eq37}),
respectively. The power and torque transferred from the plunging
region to the inner disc can be expressed as

\begin{equation}
\label{eq40} \begin{array}{l} \Delta P_{MC}^{PD} = - I_{MC}^{PD}
\varepsilon _{_{D}} \\ \quad= \left( {\frac{\Delta \varepsilon
_{_{PL}} + \Delta \varepsilon _{_{D}} }{\Delta Z_{PL} }}
\right)\left( {\frac{\Delta \Psi }{2\pi }} \right)\Omega _D =
\frac{\left( {\Omega _{PL}^m - \Omega _D } \right)\Omega _D }{\Delta
Z_{PL} }\left( {\frac{\Delta \Psi }{2\pi }} \right)^2,\end{array}
\end{equation}

\begin{equation}
\label{eq41} \Delta T_{MC}^{PD} = \frac{\Delta P_{MC}^{PD} }{\Omega
_D } = \left( {\frac{\Delta \Psi }{2\pi }} \right)^2\frac{\Omega
_{PL}^m - \Omega _D }{\Delta Z_{PL} }.
\end{equation}

In equation (\ref{eq39}) the resistance of the disc plasma is
neglected due to the perfect conductivity. The resistance of the
matter in the plunging region, $\Delta Z_{PL} $, can be determined
by the requirement of continuity of the electric current in the two
adjacent loops bounded by the innermost magnetic surface for MCHD,
i.e.,

\begin{equation}
\label{eq42} \left( {I_{MC}^{HD} } \right)_{r = r_0^ + } = \left(
{I_{MC}^{PD} } \right)_{r = r_0^ - } .
\end{equation}

Incorporating equations (\ref{eq26}), (\ref{eq27}), (\ref{eq37}),
(\ref{eq39}) and (\ref{eq42}), we have the ratio of $\Delta Z_{PL} $
to $\Delta Z_H $ as follows,

\begin{equation}
\label{eq43} \frac{\Delta Z_{PL} }{\Delta Z_H } = \frac{\Delta
\varepsilon _{PL} + \Delta \varepsilon _{_D} }{\Delta \varepsilon
_{_H} + \Delta \varepsilon _{_D} }.
\end{equation}

Considering that the innermost magnetic surface for MCHD is located
at $r_{_{PL}} = {r}_{_H} $, we infer that $\Omega _{PL} \to \Omega
_H $ due to the frame dragging of the Kerr BH. From equation
(\ref{eq43}) we have $\Delta Z_{PL} = \Delta Z_H $ with $\Delta
\varepsilon _{PL} = \Delta \varepsilon _{_H} $ for the two equal
adjacent magnetic fluxes at $r_0 $.

As a simple analysis we assume that the unknown surface resistivity
of the plasma fluid is equal to that of the BH horizon, i.e.,

\begin{equation}
\label{eq44} R_{_{PL}} = {R}_{_H} = 4\pi .
\end{equation}

\noindent Thus the resistance $\Delta Z_{PL} $ in equation
(\ref{eq39}) can be written as

\begin{equation}
\label{eq45} \begin{array}{l} dZ_{PL} = R_{_{H}} \frac{\sqrt {g_{rr}
} dr_{_{PL}} }{2\pi \varpi _{_{PL}} } \\ \quad\quad = \frac{2\tilde
r_{_{PL}}^2 d\tilde r_{_{PL}} }{\sqrt {\left( {\tilde r_{_{PL}}^4 +
a_
* ^2 \tilde r_{_{PL}}^2 + 2\tilde r_{_{PL}} a_ * ^2 } \right)\left(
{\tilde r_{_{PL}}^2 + a_ * ^2 - 2\tilde r_{_{PL}} } \right)}
}.\end{array}
\end{equation}

Finally, the total power and torque in MCHD are obtained by
integrating $\Delta P_{MC}^{HD} $ and $\Delta T_{MC}^{HD} $ over the
MCHD region as follows.


\begin{equation}
\label{eq46} \begin{array}{l} {P_{MC}^{HD} } \mathord{\left/
{\vphantom {{P_{MC}^{HD} } {P_0 }}} \right.
\kern-\nulldelimiterspace} {P_0 } = \int_{\tilde {r}_0 }^{\tilde
{r}_{_{S}} } {\left[ {\tilde {B}_D^P \left( \tilde {r} \right)}
\right]^2\left( {M\Omega _H - M\Omega _D } \right)\left( {M\Omega _D
} \right)}\\
\quad\quad\quad \times f\left( {\theta ,\tilde {r};a_ * }
\right)\left( {{ - d\tilde {r}} \mathord{\left/ {\vphantom {{ -
d\tilde {r}} {d\theta }}} \right. \kern-\nulldelimiterspace}
{d\theta }} \right)d\tilde {r} ,\end{array}
\end{equation}


\begin{equation}
\label{eq47} \begin{array}{l} {T_{MC}^{HD} } \mathord{\left/
{\vphantom {{T_{MC}^{HD} } {T_0 }}} \right.
\kern-\nulldelimiterspace} {T_0 } = \int_{\tilde {r}_0 }^{\tilde
{r}_{_{S}} } {\left[ {\tilde {B}_D^P \left( \tilde {r} \right)}
\right]^2\left( {M\Omega _H - M\Omega _D } \right)}\\
\quad\quad\quad \times f\left( {\theta ,\tilde {r};a_ * }
\right)\left( {{ - d\tilde {r}} \mathord{\left/ {\vphantom {{ -
d\tilde {r}} {d\theta }}} \right. \kern-\nulldelimiterspace}
{d\theta }} \right)d\tilde {r} ,\end{array}
\end{equation}

\noindent where the function $f\left( {\theta ,\tilde {r};a_ * }
\right)$ in equations (\ref{eq46}) and (\ref{eq47}) are give by

\begin{equation}
\label{eq48} \begin{array}{l} f\left( {\theta ,\tilde {r};a_ * }
\right) \\ \quad = \frac{2\left( {\tilde {r}^4 + a_ * ^2 \tilde
{r}^2 + 2a_
* ^2 \tilde {r}} \right)}{\left( {2\csc ^2\theta + \sqrt {1 - a_ *
^2 } - 1} \right)\left( {\tilde {r}^2 + a_ * ^2 - 2\tilde {r}}
\right)\sin \theta }.\end{array}
\end{equation}

Similarly, the total power and torque in MCPD can be calculated by
integrating $\Delta P_{MC}^{PD} $ and $\Delta T_{MC}^{PD} $ over the
MCPD region as follows.


\begin{equation}
\label{eq49} \begin{array}{l} {P_{MC}^{PD} } \mathord{\left/
{\vphantom {{P_{MC}^{PD} } {P_0 }}} \right.
\kern-\nulldelimiterspace} {P_0 } = \int_{\tilde {r}_{ms} }^{\tilde
{r}_0 } {\left[ {\tilde {B}_D^P \left( \tilde {r} \right)}
\right]^2\left( {M\Omega _{PL}^m - M\Omega _D } \right)\left(
{M\Omega
_D } \right)}\\
\quad\quad\quad \times g\left( {\tilde r_{_{PL}} ,\tilde {r};a_ * }
\right)\left( {{ - d\tilde {r}} \mathord{\left/ {\vphantom {{ -
d\tilde {r}} {d\tilde r_{_{PL}} }}} \right.
\kern-\nulldelimiterspace} {d\tilde r_{_{PL}} }} \right)d\tilde
{r},\end{array}
\end{equation}


\begin{equation}
\label{eq50} \begin{array}{l} {T_{MC}^{PD} } \mathord{\left/
{\vphantom {{T_{MC}^{PD} } {T_0 }}} \right.
\kern-\nulldelimiterspace} {T_0 } = \int_{\tilde {r}_{ms} }^{\tilde
{r}_0 } {\left[ {\tilde {B}_D^P \left( \tilde {r} \right)}
\right]^2\left( {M\Omega _{PL}^m - M\Omega _D } \right)}\\
\quad\quad\quad \times g\left( {\tilde r_{_{PL}} ,\tilde {r};a_ * }
\right)\left( {{ - d\tilde {r}} \mathord{\left/ {\vphantom {{ -
d\tilde {r}} {d\tilde r_{_{PL}} }}} \right.
\kern-\nulldelimiterspace} {d\tilde r_{_{PL}} }} \right)d\tilde
{r},\end{array}
\end{equation}

\noindent where the function $g\left( {\tilde {r}_{_{PL}} ,\tilde
{r};a_ * } \right)$ in equations (\ref{eq49}) and (\ref{eq50}) are
give by

\begin{equation}
\label{eq51} \begin{array}{l} g\left( {\tilde r_{_{PL}} ,\tilde
{r};a_
* } \right) \\ \quad = \sqrt {\left( {1 + a_
* ^2 \tilde r_{_{PL}}^{ - 2} + 2\tilde r_{_{PL}}^{ - 3} a_ * ^2 }
\right)\left( {\tilde r_{_{PL}}^2 + a_ * ^2 - 2\tilde r_{_{PL}} }
\right)} \\ \quad\quad \times \left( {\frac{\tilde {r}^4 + a_ * ^2
\tilde {r}^2 + 2a_
* ^2 \tilde {r}}{\tilde {r}^2 + a_ * ^2 - 2\tilde {r}}}
\right).\end{array}
\end{equation}

In equations (\ref{eq46})---(\ref{eq50}) we have ${ - d\tilde {r}}
\mathord{\left/ {\vphantom {{ - d\tilde {r}} {d\theta }}} \right.
\kern-\nulldelimiterspace} {d\theta } > 0$, ${ - d\tilde {r}}
\mathord{\left/ {\vphantom {{ - d\tilde {r}} {d\tilde {r}_{_{PL}}
}}} \right. \kern-\nulldelimiterspace} {d\tilde {r}_{_{PL}} } > 0$,
$P_0 \equiv B_0^2 M^2$, $T_0 \equiv B_0^2 M^3$ and

\begin{equation}
\label{eq52} \tilde {B}_D^P \left( \tilde {r} \right) \equiv {B_D^P
} \mathord{\left/ {\vphantom {{B_D^P } {B_0 }}} \right.
\kern-\nulldelimiterspace} {B_0 } = - \frac{\left( {{\Delta _D }
\mathord{\left/ {\vphantom {{\Delta _D } {A_D }}} \right.
\kern-\nulldelimiterspace} {A_D }} \right)^{1 / 2}}{2\pi
}\frac{d\tilde {\Psi }(\tilde {r},\pi \mathord{\left/ {\vphantom
{\pi {2;\tilde {r}_{ms} )}}} \right. \kern-\nulldelimiterspace}
{2;\tilde {r}_{ms} )}}{d\tilde {r}}.
\end{equation}

By using equations (\ref{eq46}), (\ref{eq47}), (\ref{eq49}) and
(\ref{eq50}) we have the curves of ${P_{MC}^{HD} } \mathord{\left/
{\vphantom {{P_{MC}^{HD} } {P_0 }}} \right.
\kern-\nulldelimiterspace} {P_0 }$, ${T_{MC}^{HD} } \mathord{\left/
{\vphantom {{T_{MC}^{HD} } {T_0 }}} \right.
\kern-\nulldelimiterspace} {T_0 }$, ${P_{MC}^{PD} } \mathord{\left/
{\vphantom {{P_{MC}^{PD} } {P_0 }}} \right.
\kern-\nulldelimiterspace} {P_0 }$ and ${T_{MC}^{PD} }
\mathord{\left/ {\vphantom {{T_{MC}^{PD} } {T_0 }}} \right.
\kern-\nulldelimiterspace} {T_0 }$ versus $a_ * $ with a given value
of $\lambda $ as shown in Figure 4.

\begin{figure}
\vspace{0.5cm}
\begin{center}
{\includegraphics[width=6cm]{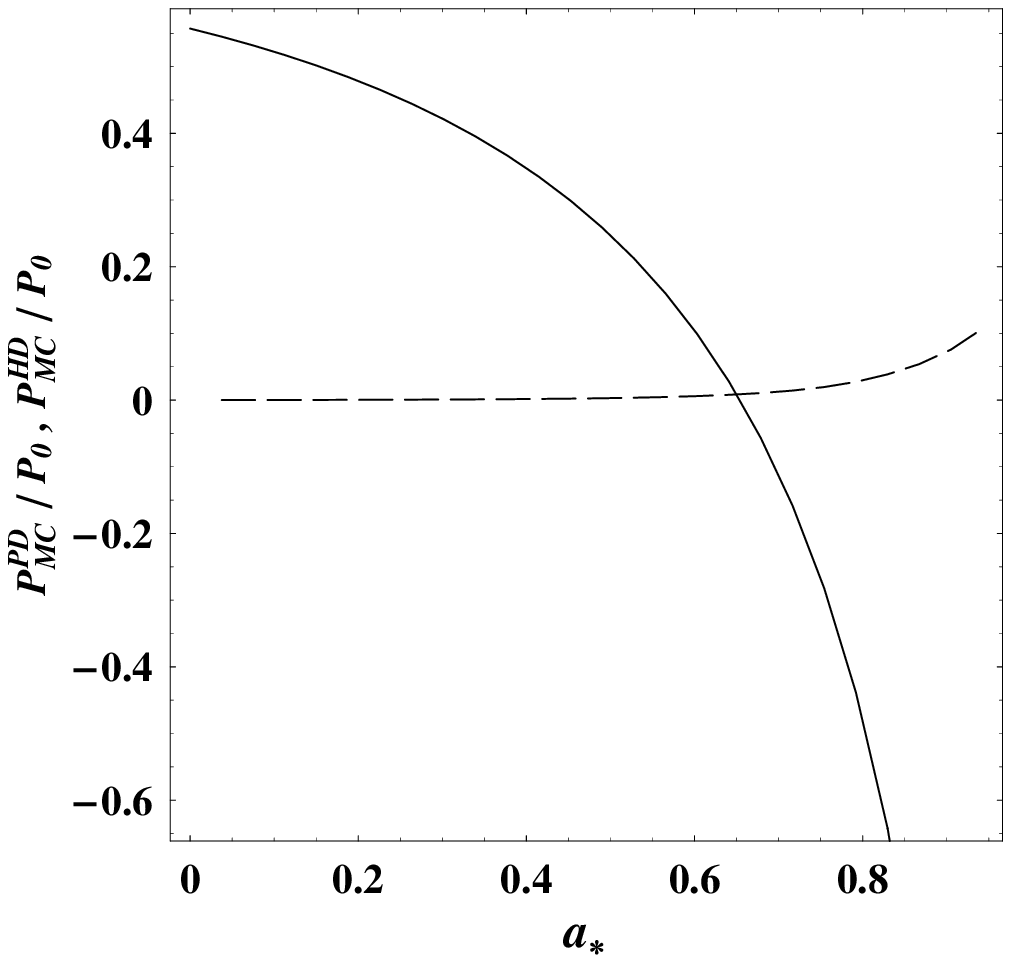}
 \centerline{(a)}
 \includegraphics[width=6cm]{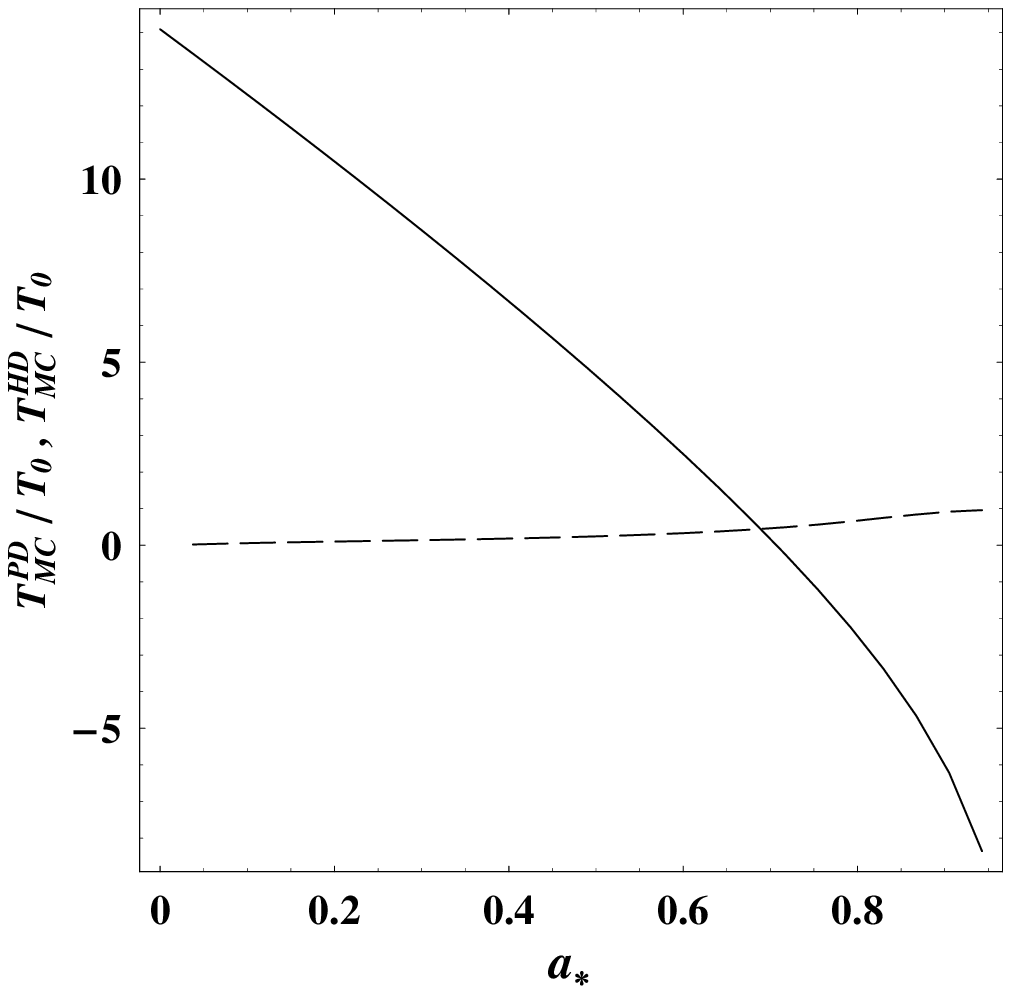}
 \centerline{(b)}}
 \caption{(a) The curves of ${P_{MC}^{PD} } / {P_0 }$
(solid line) and ${P_{MC}^{HD} } / {P_0 }$ (dashed line) versus $a_
* $; (b) the curves of ${T_{MC}^{PD} } / {T_0 }$
(solid line) and ${T_{MC}^{HD} } / {T_0 }$ (dashed line) versus $a_
* $. The parameter $\lambda = 0.5$ is assumed.}\label{fig4}
\end{center}
\end{figure}

Inspecting Figure 4, we find the following results:

(\ref{eq1}) For the increasing $a_ * $ with the given $\lambda $,
$P_{MC}^{PD} $ and $T_{MC}^{PD} $ decrease monotonically in a very
fast way, while $P_{MC}^{HD} $ and $T_{MC}^{HD} $ increase
monotonically in a much slower way.

(\ref{eq2}) The quantities $P_{MC}^{PD} $ and $T_{MC}^{PD} $ are
greater than $P_{MC}^{HD} $ and $T_{MC}^{HD} $ for low BH spins, and
the former two become negative for high BH spins, while $P_{MC}^{HD}
$ and $T_{MC}^{HD} $ remain positive for all BH spins.

Incorporating equations (\ref{eq49})---(\ref{eq51}), we have the
contours of $P_{MC}^{PD} = 0$ and ${P_{MC}^{PD} } \mathord{\left/
{\vphantom {{P_{MC}^{PD} } {P_{MC}^{HD} }}} \right.
\kern-\nulldelimiterspace} {P_{MC}^{HD} } = 1$, $T_{MC}^{PD} = 0$
and ${T_{MC}^{PD} } \mathord{\left/ {\vphantom {{T_{MC}^{PD} }
{T_{MC}^{HD} }}} \right. \kern-\nulldelimiterspace} {T_{MC}^{HD} } =
1$, in $a_ * - \lambda $ parameter space as shown in Figures 5 and
6, respectively.

\begin{figure}
\vspace{0.5cm}
\begin{center}
{\includegraphics[width=6cm]{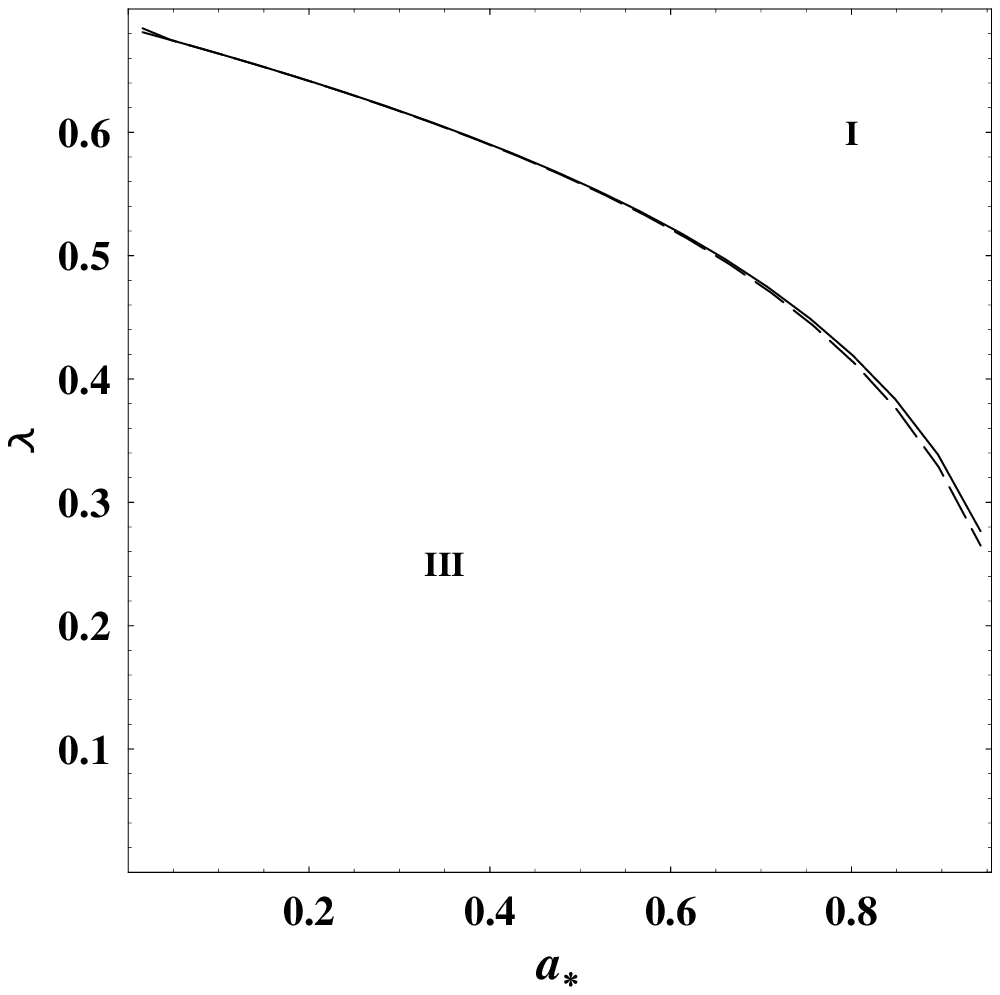}
 \centerline{(a)}
 \includegraphics[width=6cm]{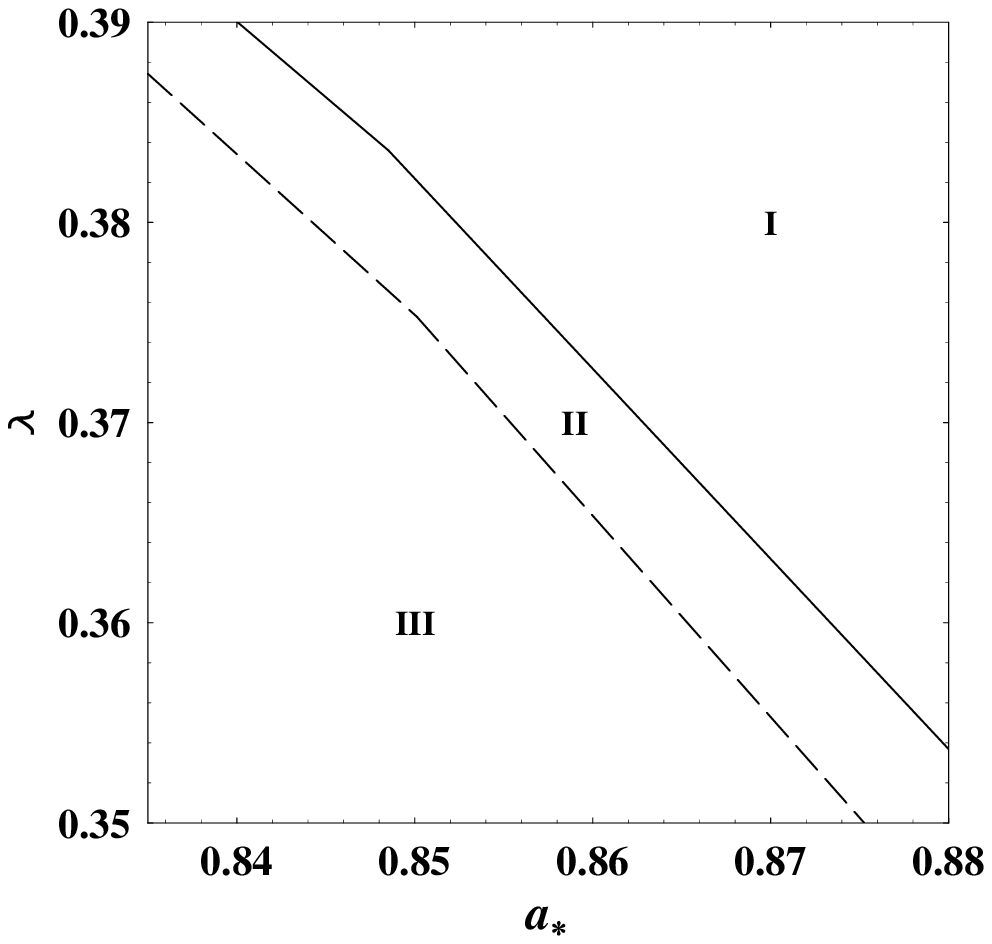}
 \centerline{(b)}}
 \caption{The contours of $P_{MC}^{PD} = 0$ (solid
line), $P_{MC}^{PD} / P_{MC}^{HD}  = 1$ (dashed line) in $a_ * -
\lambda $ parameter space for (a) $0 < a_ * < 0.9428$ and (b) $0.835
< a_ * < 0.880$. We have $P_{MC}^{PD} < 0 < P_{MC}^{HD} $, $0 <
P_{MC}^{PD} < P_{MC}^{HD} $ and $0 < P_{MC}^{HD} < P_{MC}^{PD} $ in
regions I, II and III, respectively.}\label{fig5}
\end{center}
\end{figure}

\begin{figure}
\vspace{0.5cm}
\begin{center}
{\includegraphics[width=6cm]{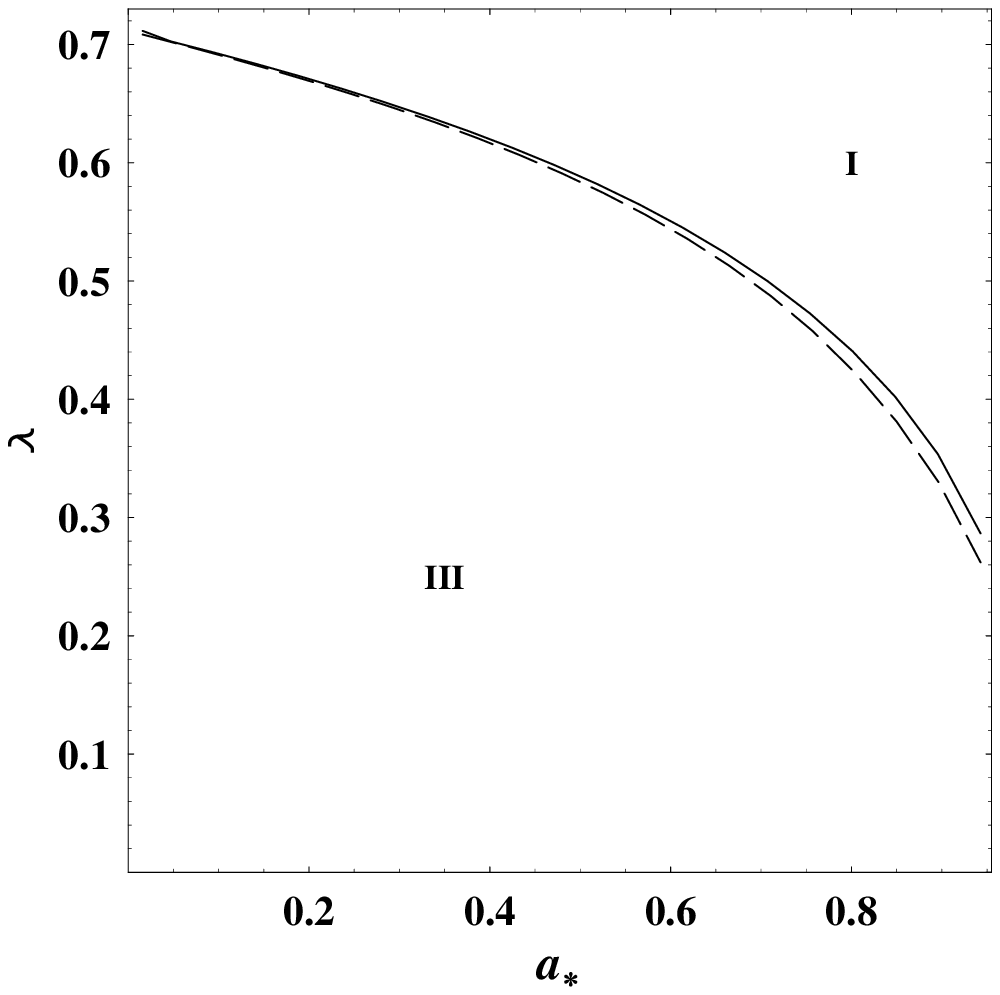}
 \centerline{(a)}
 \includegraphics[width=6cm]{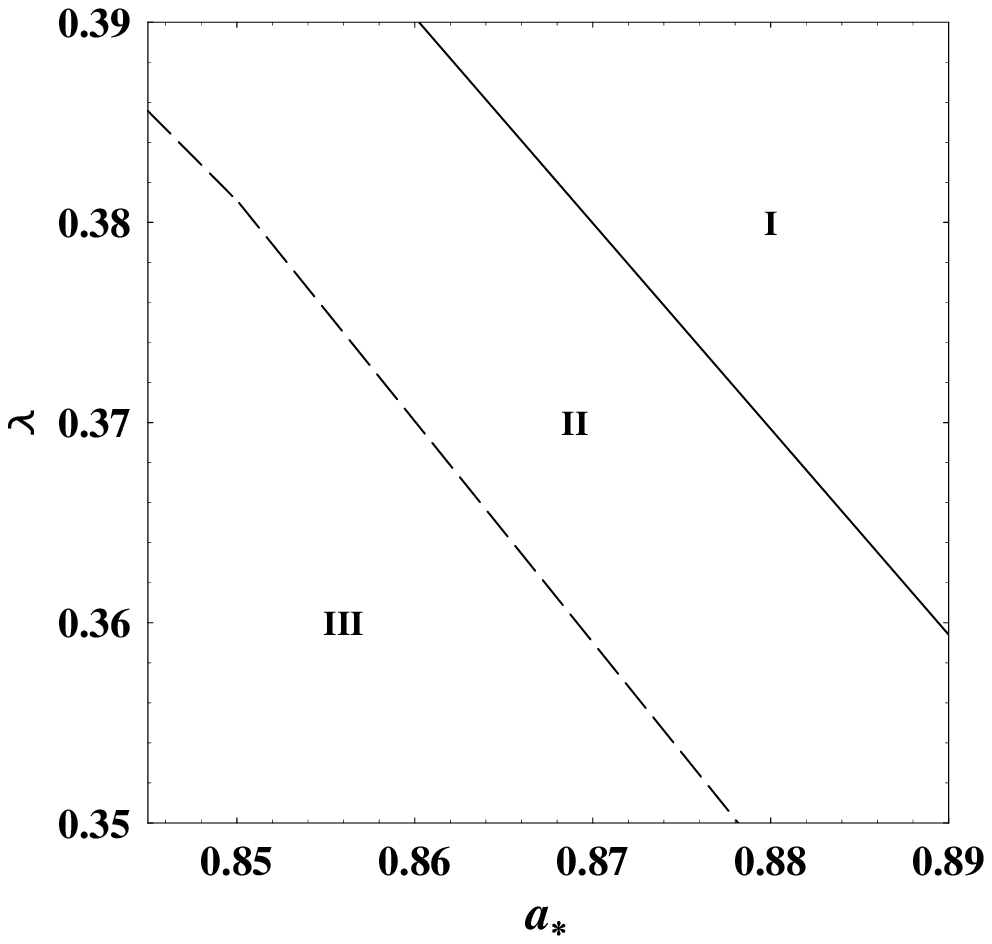}
 \centerline{(b)}}
 \caption{The contours of $T_{MC}^{PD} = 0$ (solid
line), $T_{MC}^{PD} / T_{MC}^{HD}  = 1$ (dashed line) in $a_ * -
\lambda $ parameter space for (a) $0 < a_ * < 0.9428$ and (b) $0.845
< a_ * < 0.890$. We have $T_{MC}^{PD}  < 0 < T_{MC}^{HD} $, $0 <
T_{MC}^{PD} < T_{MC}^{HD} $ and $0 < T_{MC}^{HD} < T_{MC}^{PD} $ in
regions I, II and III, respectively.}\label{fig6}
\end{center}
\end{figure}

From Figures 5 and 6 we find that the BH spin $a_ * $ and the
parameter $\lambda $ are two crucial parameters for MCPD, and it is
more effective than MCHD in transferring energy and angular momentum
into the disc for $a_* $ and $\lambda $ in the region \textbf{III.}
On the other hand, MCPD could be less effective than MCHD in this
aspect for the two parameters falling into the very narrow region
\textbf{II} as shown in Figures 6b and 7b. The integrated MCPD
transfers of energy and angular momentum are reversed for the two
parameters in region III.

\section{BH EVOLUTION AND NEGATIVE ENERGY IN PLUNGING REGION}

\subsection{BH Evolution equations }

Based on the conservation of energy and angular momentum we have the
evolution equations of a Kerr BH by considering disc accretion with
MCHD and MCPD as follows,

\begin{equation}
\label{eq53} {dM} \mathord{\left/ {\vphantom {{dM} {dt}}} \right.
\kern-\nulldelimiterspace} {dt} = E_{ms} \dot {M}_D - P_{MC}^{HD} -
P_{MC}^{PD} ,
\end{equation}

\begin{equation}
\label{eq54} {dJ} \mathord{\left/ {\vphantom {{dJ} {dt}}} \right.
\kern-\nulldelimiterspace} {dt} = L_{ms} \dot {M}_D - T_{MC}^{HD} -
T_{MC}^{PD} .
\end{equation}

Incorporating equations (\ref{eq53}) and (\ref{eq54}), we have the
evolution equation for the BH spin as follows,

\begin{equation}
\label{eq55} \begin{array}{l} {da_ * } \mathord{\left/ {\vphantom
{{da_ * } {dt}}} \right. \kern-\nulldelimiterspace} {dt} = M^{ -
2}\left( {L_{ms} \dot {M}_D - T_{MC}^{HD} - T_{MC}^{PD} } \right)
\\ \quad\quad - 2M^{ - 1}a_ * \left( {E_{ms} \dot {M}_D - P_{MC}^{HD} -
P_{MC}^{PD} } \right).\end{array}
\end{equation}

Incorporating equations (\ref{eq53})---(\ref{eq55}), we derive the
change rate of the BH entropy as follows,


\begin{equation}
\label{eq56} \begin{array}{l} T_H {dS_H } \mathord{\left/ {\vphantom
{{dS_H } {dt}}} \right. \kern-\nulldelimiterspace} {dt} = {dM}
\mathord{\left/ {\vphantom {{dM} {dt}}} \right.
\kern-\nulldelimiterspace} {dt} - \Omega _H {dJ} \mathord{\left/
{\vphantom {{dJ} {dt}}} \right. \kern-\nulldelimiterspace} {dt} =
\left( {E_{ms} - \Omega _H L_{ms} } \right)\\ \quad \times \dot
{M}_D + \left( {\Omega _H T_{MC}^{HD} - P_{MC}^{HD} } \right) +
\left( {\Omega _H T_{MC}^{PD} - P_{MC}^{PD} } \right),\end{array}
\end{equation}

\noindent where $T_H $ and $S_H $ are the temperature and entropy of
the Kerr BH, and they read (Thorne, Price {\&} MacDonald 1986)

\begin{equation}
\label{eq57} T_H = \frac{q}{4\pi M(1 + q)}, \quad S_H = 2\pi M^2(1 +
q).
\end{equation}

Considering the energy and angular momentum transferred between the
plunging region and the disc, we think that disc accretion should be
affected significantly by MCPD. There must be some relations between
the large-scale magnetic field and the accretion rate $\dot {M}_D $.
As a matter of fact these relations might be rather complicated, and
would be very different in different situations. One of them is
given by considering the balance between the pressure of the
magnetic field on the horizon and the ram pressure of the innermost
parts of an accretion flow (Moderski, Sikora {\&} Lasota 1997),
i.e.,

\begin{equation}
\label{eq58} {\left( {B_H^P } \right)^2} \mathord{\left/ {\vphantom
{{\left( {B_H^P } \right)^2} {\left( {8\pi } \right)}}} \right.
\kern-\nulldelimiterspace} {\left( {8\pi } \right)} = P_{ram} \sim
\rho c^2\sim {\dot {M}_D } \mathord{\left/ {\vphantom {{\dot {M}_D }
{\left( {4\pi {r}_{_H}^2 } \right)}}} \right.
\kern-\nulldelimiterspace} {\left( {4\pi {r}_{_H}^2 } \right)}.
\end{equation}

Considering the fact that the poloidal magnetic field arises from a
toroidal current at ISCO, we replace $B_H^P $ and ${r}_{_H} $ in
equation (\ref{eq58}) by the magnetic field near ISCO and $r_{ms} $,
respectively. In addition, we introduce a parameter $\alpha _m $ to
adjust the ratio of the magnetic pressure to the ram pressure of the
accretion flow in a thin disc, and assume the relation between the
accretion rate and the poloidal magnetic field at ISCO as follows,

\begin{equation}
\label{eq59} \dot {M}_D = {\alpha _m \left[ {B_D^p (r_{ms} )}
\right]^2r_{ms}^2 } \mathord{\left/ {\vphantom {{\alpha _m \left[
{B_D^p (r_{ms} )} \right]^2r_{ms}^2 } 2}} \right.
\kern-\nulldelimiterspace} 2 + \left( {\dot {M}_D } \right)_{MC} ,
\end{equation}

\noindent where the second term at RHS of equation (\ref{eq59}) is
the MCPD correction to the accretion rate and it reads

\begin{equation}
\label{eq60} \left( {\dot {M}_D } \right)_{MC} = - \frac{{\partial
T_{MC}^{PD} } \mathord{\left/ {\vphantom {{\partial T_{MC}^{PD} }
{\partial r}}} \right. \kern-\nulldelimiterspace} {\partial
r}}{{\partial \left( {r^2\Omega _D } \right)} \mathord{\left/
{\vphantom {{\partial \left( {r^2\Omega _D } \right)} {\partial r}}}
\right. \kern-\nulldelimiterspace} {\partial r}}.
\end{equation}

\noindent Substituting equations (\ref{eq28}) and (\ref{eq41}) into
equation (\ref{eq60}), we have

\begin{equation}
\label{eq61} \begin{array}{l} {\left( {\dot {M}_D } \right)_{MC} }
\mathord{\left/ {\vphantom {{\left( {\dot {M}_D } \right)_{MC} }
{P_0 }}} \right. \kern-\nulldelimiterspace} {P_0 } =
\\ \\ \quad \frac{2M\left[ {\tilde {B}_D^P \left( \tilde {r} \right)}
\right]^2\left( {\tilde {r}^{3 \mathord{\left/ {\vphantom {3 2}}
\right. \kern-\nulldelimiterspace} 2} + a_ * } \right)^2\left(
{\Omega _{PL}^m - \Omega _D } \right)g\left( {\tilde {r}_{_{PL}}
,\tilde {r};a_
* } \right)\left( {{d\tilde {r}} \mathord{\left/ {\vphantom
{{d\tilde {r}} {d\tilde {r}_{_{PL}} }}} \right.
\kern-\nulldelimiterspace} {d\tilde {r}_{_{PL}} }} \right)}{4\tilde
{r}a_ * + \tilde {r}^{5 \mathord{\left/ {\vphantom {5 2}} \right.
\kern-\nulldelimiterspace} 2}}.\end{array}
\end{equation}

Equations (\ref{eq60}) and (\ref{eq61}) are derived based on the
conservation of angular momentum. It is noted that the sign of
${d\tilde {r}} \mathord{\left/ {\vphantom {{d\tilde {r}} {d\tilde
{r}_{_{PL}} }}} \right. \kern-\nulldelimiterspace} {d\tilde
{r}_{_{PL}} }$ is negative, and the MCPD correction to the accretion
rate depends on both the sign and magnitude of $\Omega _{PL}^m -
\Omega _D $.

Since $\left( {\dot {M}_D } \right)_{MC} $ depends on the disc
radius, the accretion rate expressed by equation (\ref{eq59}) varies
with the disc radius, which might give rise to an unstable accretion
disc. Fortunately, the viscous force can adjust itself to compensate
for the excessive transfer of angular momentum by magnetic torque
and keep the accretion rate constant. How to constrain the
concerning parameters to ensure a stable accretion disc? We shall
deal with this issue based on the transfer of angular momentum in
the disc.

The flux of angular momentum $H$ is transferred from the plunging
region to the disc, and it is related to $T_{MC}^{PD} $ by $4\pi rH
= {\partial T_{MC}^{PD} } \mathord{\left/ {\vphantom {{\partial
T_{MC}^{PD} } {\partial r}}} \right. \kern-\nulldelimiterspace}
{\partial r}$. Based on the conservation of energy and angular
momentum Page {\&} Thorne (1974) derived the rate of angular
momentum transferred by the viscous torque in the disc as follows,

\begin{equation}
\label{eq62} g_{vis} = \frac{\left( {E^\dag - \Omega _D L^\dag }
\right)}{\left( {{ - d\Omega _D } \mathord{\left/ {\vphantom {{ -
d\Omega _D } {dr}}} \right. \kern-\nulldelimiterspace} {dr}}
\right)}f\dot {M}_D^{acc} ,
\end{equation}

\noindent where $E^\dag $ and $L^\dag $ are the specific energy and
angular momentum of the accreting particles, respectively. The
quantity $\dot {M}_D^{acc} $ can be regarded as the first term at
RHS of equation (\ref{eq59}), and the function $f = {4\pi rF}
\mathord{\left/ {\vphantom {{4\pi rF} {\dot {M}_D^{acc} }}} \right.
\kern-\nulldelimiterspace} {\dot {M}_D^{acc} }$ is related to the
radiation flux $F$ given by Page {\&} Thorne (1974).

Being required by a stationary accretion, the rate of angular
momentum transferred by the viscous torque should be no less than
that due to MCPD, i.e.,

\begin{equation}
\label{eq63} g_{vis} \ge 4\pi rH = {\partial T_{MC}^{PD} }
\mathord{\left/ {\vphantom {{\partial T_{MC}^{PD} } {\partial r}}}
\right. \kern-\nulldelimiterspace} {\partial r}.
\end{equation}

Based on equation (\ref{eq62}) and (\ref{eq63}) we have the
condition for a stationary accretion as follows,

\begin{equation}
\label{eq64} \left\{ {\begin{array}{l}
 F_{Acc} \ge 1, \\
 F_{Acc} \equiv g_{vis} \mathord{\left/ {\vphantom {g {\left( {{\partial
T_{MC}^{PD} } \mathord{\left/ {\vphantom {{\partial T_{MC}^{PD} }
{\partial r}}} \right. \kern-\nulldelimiterspace} {\partial r}}
\right)}}} \right. \kern-\nulldelimiterspace} {\left( {{\partial
T_{MC}^{PD} } \mathord{\left/ {\vphantom {{\partial T_{MC}^{PD} }
{\partial r}}} \right.
\kern-\nulldelimiterspace} {\partial r}} \right)}, \\
 \end{array}} \right.
\end{equation}

\noindent where the function $F_{Acc} $ is evaluated at $\tilde
{r}_{_{PL}} = {\left( {\tilde {r}_{ms} + \tilde {r}_{_H} } \right)}
\mathord{\left/ {\vphantom {{\left( {\tilde {r}_{ms} + \tilde
{r}_{_H} } \right)} 2}} \right. \kern-\nulldelimiterspace} 2$.

Another constraint to the accretion rate is given based on a stable
extraction of energy. The accretion rate must be great enough to
keep a stable magnetic extraction of energy from the plunging
particles. As a rough estimate, the decreasing rate of the plunging
particles' kinetic energy can be written as

\begin{equation}
\label{eq65} P_K = \frac{1}{2}\left[ {\left( {r_{_{PL}} \Omega _{PL}
} \right)^2 - \left( {r_{_{PL}} \Omega _{PL}^m } \right)^2}
\right]\dot {M}_D ,
\end{equation}

\noindent which should be no less than the MCPD power at any
position of the plunging region. Considering the fact that the
plunging particles' energy cannot be extracted completely via MCPD,
we suggest the condition for a stable energy extraction as follows,

\begin{equation}
\label{eq66} P_K \ge P_{MC}^{PD} \left( {\tilde {r}_{_{PL}} }
\right),
\end{equation}

\noindent where $P_{MC}^{PD} \left( {\tilde {r}_{_{PL}} } \right)$
is the MCPD power integrated to the position at $\tilde {r}_{_{PL}}
= {\left( {\tilde {r}_{ms} + \tilde {r}_{_H} } \right)}
\mathord{\left/ {\vphantom {{\left( {\tilde {r}_{ms} + \tilde
{r}_{_H} } \right)} 2}} \right. \kern-\nulldelimiterspace} 2$. Thus
the condition (\ref{eq66}) can be rewritten as

\begin{equation}
\label{eq67} F_{Energy} \equiv P_K - P_{MC}^{PD} \left( {\tilde
{r}_{_{PL}} } \right) \ge 0.
\end{equation}

\subsection{Condition for Negative Energy of Plunging Particles }

Not long ago, L00 discussed the possibility of negative energy
carried by accreting particles onto a BH due to the magnetic
extraction of energy to the remote loads. An interesting issue is
whether negative energy can be realized via MCPD. We are going to
discuss this possibility with the constraints based on several
physical considerations.

Following L00, we have the condition for the negative energy of the
plunging particles as follows,

\begin{equation}
\label{eq68} P_{PL} = \dot {M}_D E_{ms} - P_{MC}^{PD} ,
\end{equation}

\noindent where $P_{PL} $ is the rate of the net energy brought into
the BH by the plunging particles. The negative energy is determined
by $P_{PL} < 0$ with the following constraints.

(\ref{eq1}) The rate of the BH entropy given by equation
(\ref{eq56}) is not negative, which implies that the BH entropy can
never decrease required by the second law of BH thermodynamics
(Thorne, Price {\&} MacDonald 1986);

(\ref{eq2}) The modified accretion rate expressed by equation
(\ref{eq59}) remains positive to provide the plunging particles in a
continuous way;

(\ref{eq3}) The stationary accretion is guaranteed by equation
(\ref{eq64});

(\ref{eq4}) A stable energy extraction from the plunging particles
is ensured based on equation (\ref{eq67}).

Incorporating equations (\ref{eq56})---(\ref{eq68}), we have the
contours of ${dS_H } \mathord{\left/ {\vphantom {{dS_H } {dt}}}
\right. \kern-\nulldelimiterspace} {dt} = 0$, $\dot {M}_D = 0$,
$F_{Acc} = 1$, $F_{Energy} = 0$ and $P_{PL} = 0$ in the $a_ * -
\lambda $ parameter space as shown in Figure 7.

\begin{figure}
\vspace{0.5cm}
\begin{center}
{\includegraphics[width=6cm]{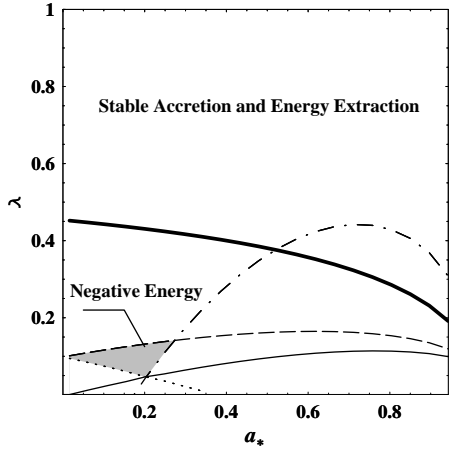}
 \centerline{(a)}
 \includegraphics[width=6cm]{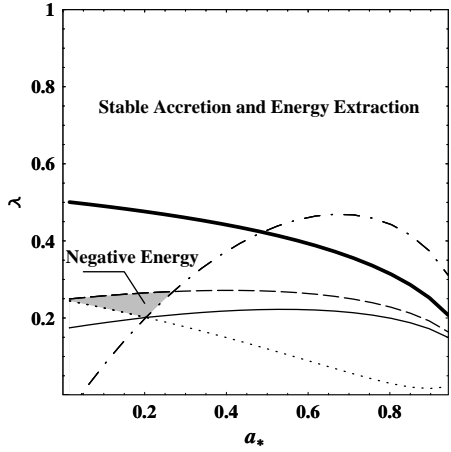}
 \centerline{(b)}}
 \caption{The contours of $P_{PL} = 0$ (dashed line), $dS_H / dt = 0$ (dotted line), $\dot {M}_D = 0$
(solid line), $F_{Acc} = 1$ (dot-dashed line) and $F_{Energy} = 0$
(thick solid line) in $a_ * - \lambda $ parameter space with $0 < a_
* < 0.9428$ and $0 < \lambda < 1$ for (a) $\alpha _m = 0.039$ and
(b) $\alpha _m = 0.030$.}\label{fig7}
\end{center}
\end{figure}

In Figure 7 the shaded region indicated "Negative Energy" is bounded
by four contours, i.e. $P_{PL} \leq 0$, $dS_H / dt \geq 0$, $\dot
{M}_D \geq 0$ and $F_{Acc} \geq 1$ in dashed, dotted, solid
dot-dashed lines, respectively. However, the negative energy is
excluded by the contour of $F_{Energy} \geq 0$ in thick solid line,
above which the stable energy extraction from the plunging particles
is required. The region above the thick solid line and the
dot-dashed line is indicated "Stable Accretion and Energy
Extraction", in which energy extraction via MCPD can work in a
stable way.

Thus we conclude, although the negative energy can be delivered to
the BH via MCPD with the validity of the second law of the BH
thermodynamics, it cannot be realized in a stable way in our model.

\section{EFFECTS OF MCPD AND MCHD ON EFFICIENCY OF ENERGY RELEASE}

Although MCHD and MCPD are related to the magnetic field
configuration supported by a toroidal current flowing on the
equatorial plane of the Kerr BH, these two mechanisms are different
in several aspects.

(\ref{eq1}) Energy and angular momentum are extracted and
transferred from the Kerr BH to the disc by virtue of MCHD, while
those are extracted and transferred from the plunging particles to
the inner disc by virtue of MCPD.

(\ref{eq2}) MCHD can work in a suspended accretion state as
suggested by van Putten {\&} Ostriker (2001), while MCPD cannot
operate without continuous replenishment of accreting particles.

Therefore a stable accretion and energy extraction are indispensable
in the model containing the coexistence of MCHD and MCPD. Based on
the BH evolution equation (\ref{eq53}), we can easily obtain the
efficiencies of releasing energy as follows.

\begin{equation}
\label{eq69} \eta = 1 - {\left( {{dM} \mathord{\left/ {\vphantom
{{dM} {dt}}} \right. \kern-\nulldelimiterspace} {dt}} \right)}
\mathord{\left/ {\vphantom {{\left( {{dM} \mathord{\left/ {\vphantom
{{dM} {dt}}} \right. \kern-\nulldelimiterspace} {dt}} \right)} {\dot
{M}_D }}} \right. \kern-\nulldelimiterspace} {\dot {M}_D } = \eta
_{_DA} + \eta _{MC}^{HD} + \eta _{MC}^{PD} ,
\end{equation}

\noindent where $\eta _{_DA} $, $\eta _{MC}^{HD} $ and $\eta
_{MC}^{PD} $ are respectively the efficiencies due to disc
accretion, MCHD and MCPD, converting the rest energy of the
accreting particles into the radiation energy. These efficiencies
are expressed as

\begin{equation}
\label{eq70}  \left\{ {\begin{array}{l} \eta _{_DA} = 1 - E_{ms} , \\
\eta _{MC}^{HD} = {P_{MC}^{HD} } \mathord{\left/ {\vphantom
{{P_{MC}^{HD} } {\dot {M}_D }}} \right. \kern-\nulldelimiterspace}
{\dot {M}_D
},\\
\eta _{MC}^{PD} = {P_{MC}^{PD} } \mathord{\left/ {\vphantom
{{P_{MC}^{PD} } {\dot {M}_D }}} \right. \kern-\nulldelimiterspace}
{\dot {M}_D }.  \end{array}}\right.
\end{equation}

Incorporating equation (\ref{eq70}) with the concerning expressions
for $E_{ms} $, $\dot {M}_D $, $P_{MC}^{HD} $ and $P_{MC}^{PD} $, we
have the curves of $\eta _{_DA} $, $\eta _{MC}^{PD} $ and $\eta
_{MC}^{HD} $ versus $a_ * $ for the different values of $\alpha _m $
and $\lambda $ as shown in Figure 8.

\begin{figure}
\vspace{0.5cm}
\begin{center}
{\includegraphics[width=6cm]{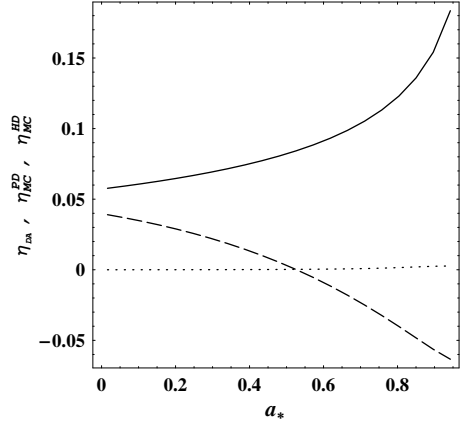}
 \centerline{(a)}
 \includegraphics[width=6cm]{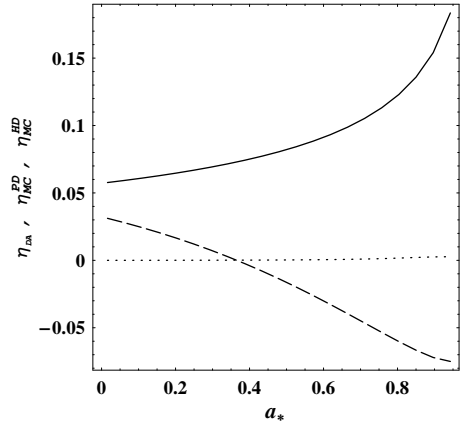}
 \centerline{(b)}}
 \caption{The curves of $\eta _{_DA} $, $\eta _{MC}^{PD}
$ and $\eta _{MC}^{HD} $ versus $a_ * $ respectively in solid,
dashed and dotted lines for (a) $\alpha _m = 0.039$, $\lambda =
0.55$ and (b) $\alpha _m = 0.03$, $\lambda = 0.6$.}\label{fig8}
\end{center}
\end{figure}

The efficiency $\eta _{MC}^{PD} $ is greater (less) than $\eta
_{MC}^{HD} $ for the lower (higher) BH spin, while it is generally
less than $\eta _{_DA} $ as shown in Figure 8. In order to compare
the efficiencies $\eta _{MC}^{PD} $ and $\eta _{_DA} $ in an
accurate way we have the contours of ${\eta _{MC}^{PD} }
\mathord{\left/ {\vphantom {{\eta _{MC}^{PD} } {\eta _{_DA} }}}
\right. \kern-\nulldelimiterspace} {\eta _{_DA} } = 1$, $F_{Acc} =
1$ and $F_{Energy} = 0$ in $a_ * - \lambda $ parameter space as
shown in Figure 9.

\begin{figure}
\vspace{0.5cm}
\begin{center}
{\includegraphics[width=6cm]{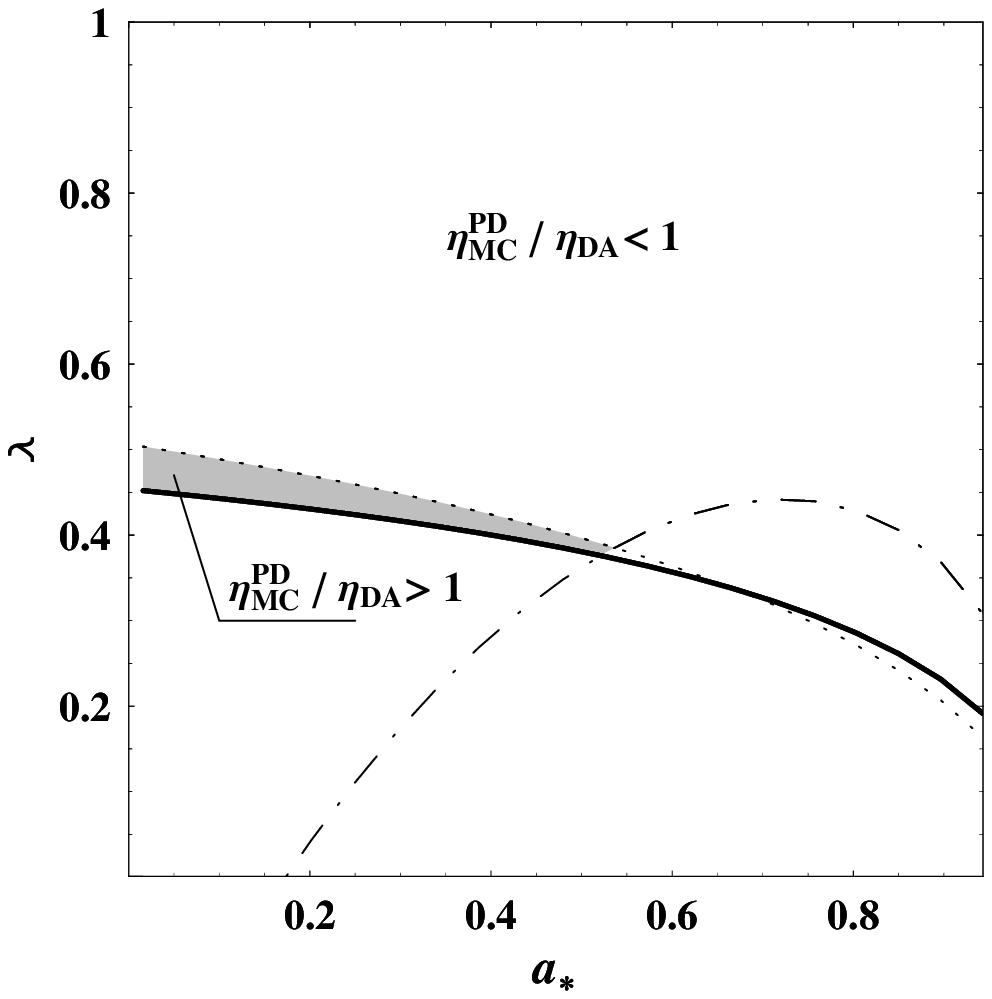}
 \centerline{(a)}
 \includegraphics[width=6cm]{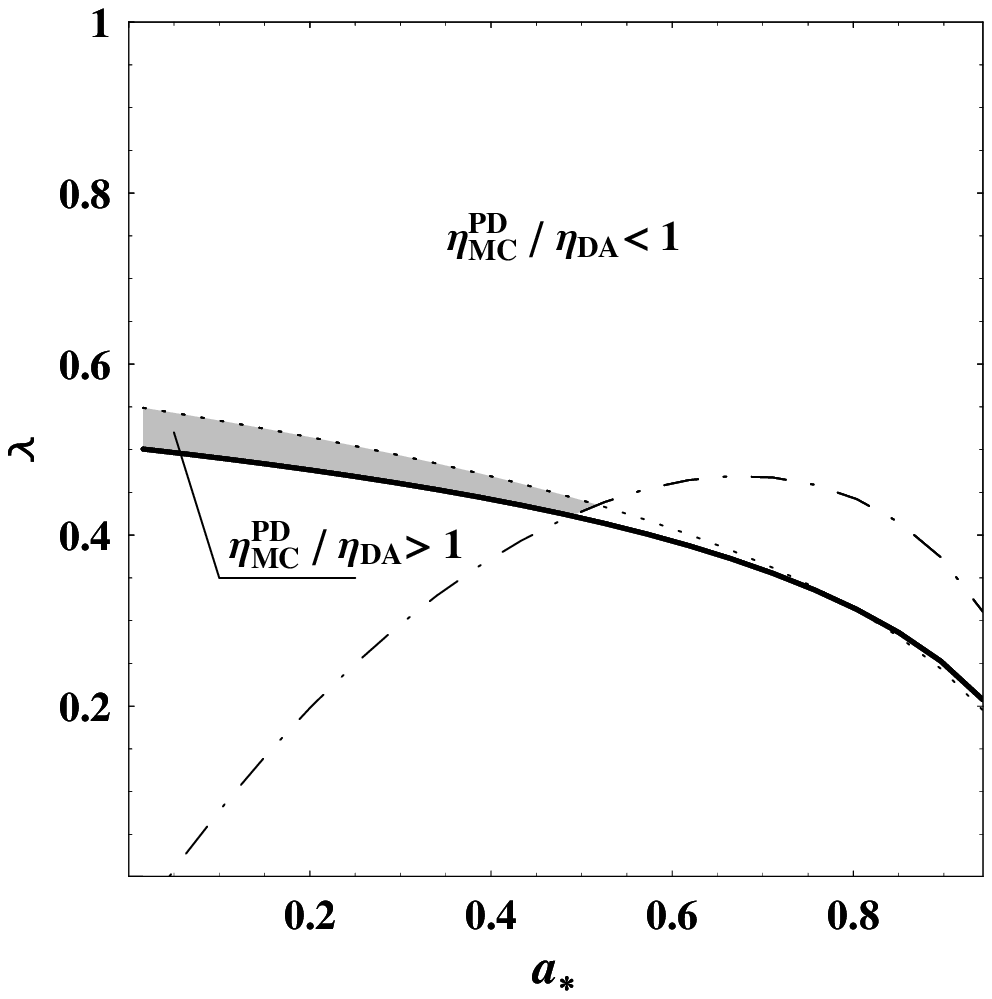}
 \centerline{(b)}}
 \caption{Contours in $a_ * - \lambda $ parameter
space of $\eta _{MC}^{PD} / \eta _{_DA}  = 1$, $F_{Energy} = 0$ and
$F_{Acc} = 1$ in dotted, solid and dot-dashed lines, respectively.
The parameter $\alpha _m $=0.039 and 0.03 are given in Figures 10a
and 10b, respectively.}\label{fig9}
\end{center}
\end{figure}

Inspecting Figure 9, we find that efficiency $\eta _{MC}^{PD} $ is
generally less than $\eta _{_DA} $ in a wide value range of $a_ * $
and $\lambda $ for the given values of $\alpha _m $, and the
possibility of $\eta _{MC}^{PD} > \eta _{_DA} $ is confined in a
very narrow shaded region indicated by ``${\eta _{MC}^{PD} }
\mathord{\left/ {\vphantom {{\eta _{MC}^{PD} } {\eta _{_DA} }}}
\right. \kern-\nulldelimiterspace} {\eta _{_DA} } > 1$'' in Figure
9.

\section{DISCUSSION}

In this toy model the magnetic field configuration is generated by a
single electric current flowing at ISCO of a thin disc, and we
regard this as a starting point for dealing with the magnetic field
configuration generated by the electric current distributed in more
realistic way around a BH. We concentrate the discussion on the
magnetic extraction of energy and angular momentum from the plunging
particles, and we describe our model phenomenally by using two
parameters instead of resolving the extremely complicated MHD
equations within ISCO. The parameter $\lambda $ is used to describe
the angular velocity of the plunging particles, and the parameter
$\alpha _m $ is invoked to adjust the accretion rate affected by
MCPD. It is noted that the parameters $a_ * $, $\lambda $ and
$\alpha _m $ are crucial for a stable MCPD process, although the
MCPD power and torque are independent of the parameter $\alpha _m $
according to equations (\ref{eq49})--(\ref{eq51}).

The key point in this model is how to constrain these parameters
based on some reasonable considerations. In this model four
requirements are given: (i) a positive accretion rate, (ii) an
increasing BH entropy, (iii) a stationary accretion and (iv) a
stable energy extraction for the plunging particles. Required by
these constraints the following results are obtained in our model.

(\ref{eq1}) MCPD could be more important than MCHD in transferring
energy and angular momentum to the disc. As shown in Figures 4--6,
the power and torque in MCPD are stronger than those in MCHD for low
BH spins.

(\ref{eq2}) Although negative energy can be delivered to the BH by
the plunging particles with the increasing BH entropy it cannot be
realized via MCPD in a stable way as shown in Figure 7.

The efficiency of releasing energy due to MCPD could be greater than
that due to MCHD for low BH spins, while it is generally less than
the efficiency of disc accretion for a wide range of $a_ * $ and
$\lambda $ as shown in Figure 9.

It is assumed in $\S$ 2.2 that the toroidal current located at
$r_{ms} $ has a very small circular section of radius
${r}'_\varepsilon = \varepsilon r_{ms} $, and $\varepsilon = 10^{ -
4}$ is taken in calculations. The parameter $\varepsilon $ is
introduced to avoid an infinite magnetic field as $r $close to the
toroidal current. The magnetic field outside ${r}'_\varepsilon =
\varepsilon r_{ms} $ are not affected by the value of $\varepsilon
$, while the contribution of the magnetic field to the MCPD effects
can be neglected if $\varepsilon $ is small enough. Thus the
influences of $\varepsilon $ on the MCPD effects are very little.
For example, required by ${P_{MC}^{PD} } \mathord{\left/ {\vphantom
{{P_{MC}^{PD} } {P_{MC}^{HD} }}} \right. \kern-\nulldelimiterspace}
{P_{MC}^{HD} } = 1$with $a_\ast $=0.294, we have $\lambda
$=0.6187569089, 0.6187569092 and 0.6187569342 for $\varepsilon $=
10$^{ - 5}$, 10$^{ - 4}$ and 10$^{ - 3}$, respectively. The
variation of $\lambda $ arising from that of $\varepsilon $ is
${\Delta \lambda } \mathord{\left/ {\vphantom {{\Delta \lambda }
{\Delta \varepsilon }}} \right. \kern-\nulldelimiterspace} {\Delta
\varepsilon } < 3\times 10^{ - 5}$.

Another issue related to our model lies in prescription
(\ref{eq44}), in which the surface resistivity of the plunging
region is assumed to be $R_{{PL}} = {R}_{H} = 4\pi $. The
resistivity ${R}_{H} $ is defined such that the electric field at
the BH horizon is equal in magnitude to the magnetic field. We can
show that the prescription (\ref{eq44}) does not lead to the result
that the electric field at the surface of the plunging region is
greater than the magnetic field there.

The toroidal component of the magnetic field at the surface of the
disc is expressed by equation (\ref{eq25}). Similarly, we can derive
the expressions for the toroidal magnetic field at the surface of
the plunging region as follows,

\begin{equation}
\label{eq71} B_{PL}^T = {2I_{MC}^{PD} } \mathord{\left/ {\vphantom
{{2I_{MC}^{PD} } {\left( {\varpi _{PL} \alpha } \right)_{\theta =
\pi \mathord{\left/ {\vphantom {\pi 2}} \right.
\kern-\nulldelimiterspace} 2} }}} \right. \kern-\nulldelimiterspace}
{\left( {\varpi _{PL} \alpha } \right)_{\theta = \pi \mathord{\left/
{\vphantom {\pi 2}} \right. \kern-\nulldelimiterspace} 2} }.
\end{equation}

The toroidal electric field is zero in a stationary axisymmetric
magnetosphere (MT82), while the poloidal electric field at the
surface of the plunging region is related to $I_{MC}^{PD} $ by Ohm's
law, i.e.,

\begin{equation}
\label{eq72} \alpha E_{PL} = R_{{PL}} {I_{MC}^{PD} } \mathord{\left/
{\vphantom {{I_{MC}^{PD} } {2\pi \varpi _{PL} }}} \right.
\kern-\nulldelimiterspace} {2\pi \varpi _{PL} },
\end{equation}

\noindent where $\alpha E_{PL} $ is the electric field converted to
a `per unit globe time' basis (MT82). Incorporating equations
(\ref{eq25}) and (\ref{eq72}) we have

\begin{equation}
\label{eq73} {E_{PL} } \mathord{\left/ {\vphantom {{E_{PL} }
{B_{PL}^T }}} \right. \kern-\nulldelimiterspace} {B_{PL}^T } =
{R_{{PL}} } \mathord{\left/ {\vphantom {{R_{{PL}} } {4\pi }}}
\right. \kern-\nulldelimiterspace} {4\pi }.
\end{equation}

\noindent Thus the prescription (\ref{eq44}) leads to $E_{PL} =
B_{PL}^T $ and $E_{PL} < B_{PL} = \sqrt {\left( {B_{PL}^T }
\right)^2 + \left( {B_{PL}^P } \right)^2} $.

In this paper we do not fit the very steep emissivity observed in
some BH systems, such as galaxy MCG-6-30-15. In fact, a very steep
emissivity can be also produced by virtue of MCPD, since the energy
and angular momentum transferred from the plunging region are very
concentrated in the inner disc. We shall work out a more realistic
model to relate MCPD to the observations in our future work.

\section*{Acknowledgements}

This work is supported by the National Natural Science Foundation of
China under Grant Numbers 10373006, 10573006 and 10121503. We are
very grateful to the anonymous referee for his (her) instructive
comments and a lot helpful suggestions on our previous manuscript.

\end{document}